\documentclass[pdflatex,sn-mathphys-num]{sn-jnl}

\usepackage{microtype}
\usepackage{graphicx}%
\usepackage{multirow}%
\usepackage{amsmath,amssymb,amsfonts}%
\usepackage{amsthm}%
\usepackage{mathrsfs}%
\usepackage[title]{appendix}%
\usepackage{xcolor}%
\usepackage{textcomp}%
\usepackage{manyfoot}%
\usepackage{booktabs}%
\usepackage{algorithm}%
\usepackage{algorithmicx}%
\usepackage{algpseudocode}%
\usepackage{listings}%
\usepackage{enumitem}
\usepackage{array}
\renewcommand{\arraystretch}{1.4}
\usepackage{longtable}
\usepackage{url}

\let\orcidlogo\relax 
\usepackage{orcidlink}
\theoremstyle{thmstyleone}%
\theoremstyle{thmstyletwo}%

\theoremstyle{thmstylethree}%

\usepackage{comment}

\begin{document}

\title[fNIRS in Game-Integrated Learning]{Neurophysiological Insights into Multimedia-based Education: A PRISMA-ScR Review of fNIRS in Game-Integrated Learning Systems}


\author{\fnm{Shayla} \sur{Sharmin}\,\orcidlink{0000-0001-5137-1301}}
\email{shayla@udel.edu}

\author{\fnm{Gael } \sur{Lucero-Palacios}\, \orcidlink{0009-0005-0166-4751}}\email{gael@udel.edu}

\author{\fnm{Behdokht} \sur{ Kiafar}\, \orcidlink{0009-0001-4415-1332}}\email{kiafar@udel.edu}

\author{\fnm{Md Fahim} \sur{  Abrar}\, \orcidlink{0009-0009-5157-7807}}\email{fahim@udel.edu}

\author{\fnm{Mohammad } \sur{ Al-Ratrout}\, \orcidlink{0009-0002-0180-4345}}\email{mratrout@udel.edu}

\author{\fnm{Aditya} \sur{ Raikwar}\, \orcidlink{0009-0005-9427-9253}}\email{adirar@udel.edu}

\author{\fnm{Roghayeh Leila} \sur{ Barmaki}\,\orcidlink{0000-0002-7570-5270}}\email{rlb@udel.edu}
\affil{\orgdiv{Computer and Information Sciences}, \orgname{University of Delaware}, \orgaddress{ \city{Newark}, \postcode{19716}, \state{DE}, \country{USA}}}


\abstract{ 

Game-integrated learning systems (GILS) are a growing form of multimedia education. Neurophysiological evidence can complement behavioral and learning measures when examining how users interact with these environments. This scoping review follows PRISMA-ScR and synthesizes 20 empirical studies published between 2014 and 2025, identified through searches updated in April 2026, in which functional near-infrared spectroscopy (fNIRS) measured task-related cortical hemodynamic responses during GILS use. The review maps learning contexts, multimedia platforms, outcome measures, and methodological approaches across the included studies. Individual studies reported differences in cortical responses across interaction conditions and associations with behavioral or learning outcomes. However, only 10 of the 20 studies assessed learning outcomes, and methodological heterogeneity limits cross-study comparisons. These findings do not establish that specific multimedia platforms cause particular neural or learning outcomes. Future research should prioritize standardized methods, ecologically valid learning settings, and evaluation of the feasibility and educational value of real-time neural adaptation.}




\maketitle

\section{Introduction}\label{sec:introduction} 
 

Game-integrated learning systems (GILS) are now used across a wide range 
of multimedia formats, from screen-based educational games to immersive 
virtual reality (VR) and augmented reality (AR) simulations and 
multisensory devices. These platforms give designers new ways to support 
learning and open new paths to examine learning with multimedia 
technologies. 
Different platforms provide different forms of interaction, and examining 
task-related neurophysiological responses across these learning contexts 
may provide complementary information about how learners interact with 
multimedia environments.

Traditional evaluation methods, such as tests, surveys, and observations, 
can measure what learners know or how they perform. 
Adding neurophysiological data alongside these methods can provide 
complementary information about physiological responses during learning 
\citep{devidesco2021Neuroscience,dingler2017building,varma2008scientific}. 
Mayer's Cognitive Theory of Multimedia Learning (CTML) provides a theoretical 
basis for considering how multimedia information is processed 
\citep{mayer2014incorporating}. Behavioral, learning, subjective, and 
neurophysiological measures can therefore provide complementary 
perspectives for examining multimedia learning environments.

 
\subsection{Conceptual Background: Game-Integrated Learning and fNIRS} 


The GILS literature uses related terms such as game-based learning, 
educational games, and serious games \citep{becker2021s}. In this review, 
these terms represent learning contexts in which gameplay or game-like 
interaction is used for educational, training, or skill-development 
purposes. We also distinguish the learning context from the multimedia 
delivery platform through which the activity is presented. These 
platforms range from conventional screen-based systems to immersive 
VR/AR environments and specialized interactive systems. The operational categories used to classify the included studies and representative examples of each category are described in \autoref{sec: Classification}.

Among neurophysiological methods, functional near-infrared spectroscopy 
(fNIRS) has emerged as 
a potentially useful tool
for studying game-integrated learning systems. It is a portable, 
non-invasive brain-imaging tool that measures changes in blood 
oxygenation in the cortex as an indirect measure of brain activity 
\citep{lloyd2010illuminating,ning2024fnirs,gao2024advances}. 
Because it permits greater mobility than some other neuroimaging 
approaches and is less sensitive to electrical noise from screens and 
game devices, it can be used in interactive learning environments where 
methods such as functional magnetic resonance imaging (fMRI) may face 
practical constraints 
\citep{LAMB201814,ninaus2014neurophysiological}. 
fNIRS does not directly measure learning, engagement, mental effort, or 
cognitive load. Rather, it measures task-related cortical hemodynamic 
responses, which may be interpreted in relation to cognitive processes 
when considered alongside the task context and complementary behavioral, 
learning, or subjective measures.

\subsection{Prior Reviews, Research Gap, and Motivation} 
Previous reviews have approached neurophysiological research in education and learning from several complementary perspectives. \citet{ninaus2014neurophysiological} reviewed neurophysiological methods, including EEG, fNIRS, and fMRI, in serious games and virtual environments, highlighting the heterogeneity of study designs, game genres, and research questions in this area. \citet{tinga2020non} considered non-invasive neurophysiological measures broadly across learning and training, whereas \citet{tenorio2022brain} systematically reviewed the use of multiple brain-imaging techniques in technology-enhanced learning. More 
specifically for fNIRS, ~\citet{Zhan2023Applying} reviewed 
its application across educational research, including the populations, 
educational domains, and research contexts represented in this 
literature. More recently, ~\citet{Nouri2025Scoping} 
examined educational neurotechnology broadly, including fNIRS alongside 
EEG, MRI/fMRI, and other neurotechnologies. From the perspective of 
immersive learning, ~\citet{Zhou2026Physiological} reviewed 
physiological sensing and AI-based signal analysis in immersive VR 
learning, with attention to acquisition, preprocessing, modeling, 
validation, and readiness for real-time deployment. 

Together, these reviews establish a broader foundation for 
neurophysiological measurement in education, technology-enhanced 
learning, and immersive learning environments. Their primary scopes, 
however, differ from the specific intersection examined in the present 
review. Here, we focus specifically on fNIRS studies conducted in 
game-integrated learning contexts and map the learning and multimedia 
settings, interaction modalities, behavioral and learning outcomes, 
study designs, fNIRS instrumentation and cortical coverage, and 
analytical approaches represented in this literature. Rather than 
treating differences observed across separate studies as effects of the 
multimedia platforms themselves, we characterize how fNIRS has been 
used across different GILS contexts and identify recurring 
methodological patterns and gaps. 
 
This narrower intersection motivates examining the fNIRS--GILS 
literature as a distinct body of evidence and considering what the 
current evidence can, and cannot, support regarding the use of fNIRS 
in future multimedia and neuroadaptive learning systems. 

\subsection{Review Scope, Objectives, and Research Questions}
To address this gap, we conducted a scoping review following the PRISMA-ScR framework \citep{prisma_scoping_review}, with searches updated through April 2026. We identified 1,563 records and included 20 empirical studies published between 2014 and 2025.
This review maps how fNIRS has been used in GILS contexts, examines reported relationships among task-related cortical hemodynamic responses, behavioral measures, and learning outcomes, and identifies methodological gaps and directions for future research. 

\begin{figure}[t] 
\includegraphics[width=\textwidth]{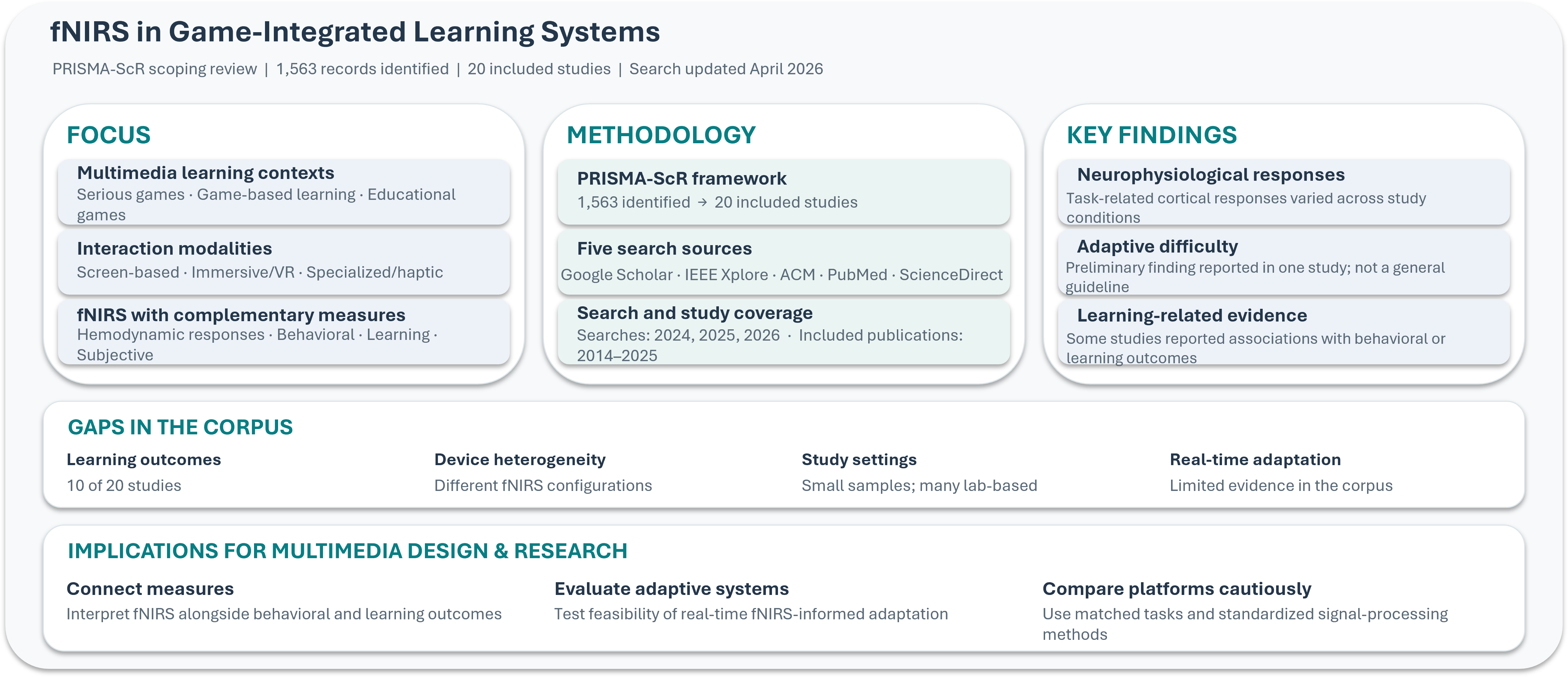} 
\caption{Scoping review overview of fNIRS in game-integrated learning 
systems (GILS): learning and multimedia contexts, fNIRS-based 
neurophysiological measurement, methodology (PRISMA-ScR framework, 
1,563 records screened, 20 studies included), methodological gaps in 
the corpus, and implications for future multimedia learning research.} 
\label{fig:teaser} 
\end{figure}

A scoping-review approach was selected because this literature is relatively small and methodologically heterogeneous, spanning different learning contexts, study designs, outcome measures, fNIRS systems, cortical regions, and analytical procedures. Accordingly, the review was designed to map the breadth and characteristics of the available evidence rather than to estimate a pooled intervention effect, consistent with the evidence-mapping purpose of a scoping review. 
~\autoref{fig:teaser} provides an overview of the review scope, 
linking game-integrated learning contexts, fNIRS-based 
neurophysiological measurement, and the outcome categories examined 
in the included studies. 
We structure this review around two research questions: 
 
\begin{enumerate}[label=$RQ_{\arabic*}$] 
\item {
How have fNIRS-measured task-related cortical hemodynamic responses 
been examined across multimedia contexts in GILS, and what relationships 
have been reported among neurophysiological, behavioral, and learning 
outcomes? 
} 
\item { 
What methodological and ecological gaps in existing fNIRS-GILS research 
limit the development of real-time adaptive multimedia learning systems, 
and what directions are needed to bridge this gap? 
} 
 
\end{enumerate}

The paper is structured as follows: \autoref{sec:methodology} describes the search strategy, eligibility criteria, screening procedure, and data extraction and verification process. Section 3 presents the review results. \autoref{sec:selection of sources of evidence} reports the selection of evidence sources and the PRISMA-ScR flow, while \autoref{sec: Classification} presents the characteristics and classification of the included studies, including learning platforms, multimedia delivery platforms, outcome measures, and study types. \autoref{sec:high level} examines corpus-level trends and methodological patterns, and \autoref{sec:detailed} provides a detailed synthesis of findings from comparative and cognitive response studies. Finally, \autoref{sec:Discussion} addresses the research questions through cross-study interpretation, discusses methodological and ecological limitations, and outlines future research directions. We conclude with an overall summary in \autoref{sec:Conclusion}.
\section{Methodology}\label{sec:methodology}

\subsection{Search Method}
We followed the PRISMA-ScR \cite{prisma_scoping_review}, a framework for conducting this review process for searching and screening methods. The paper search was initially conducted from June 13 to 16, 2024, and updated on June 2, 2025. 
  A final search update was conducted on April 1, 2026, before manuscript submission to ensure the review reflects the most current state of the literature. All three search phases followed identical procedures, with the same team applying the same inclusion and exclusion criteria throughout.  The screening process was managed using a shared Excel sheet. Reviewers recorded their initial screening decisions before disagreements were discussed and resolved. The papers were collected from five digital libraries: Google Scholar, Institute of Electrical and Electronics Engineers (IEEE), Association for Computing Machinery (ACM), ScienceDirect (SD), and PubMed. 


For Google Scholar, each search query was manually reviewed across all result pages made available by Google Scholar at the time of the search, rather than stopping at a predefined page or result cutoff. No automated scraping was used. Records were manually saved to the Google Scholar Library using the platform’s ``Save'' function and subsequently exported from the Library as CSV files for screening. Duplicate records and other ineligible records were identified and removed during the subsequent screening process.

\subsection{Search Criteria} 

We conducted a multi-stage search strategy without time constraints to cover a broad range of the available literature. The earliest included study is from 2014, reflecting the period in which fNIRS began to appear in game-integrated learning systems within our search results. Initially, structured queries using Boolean operators (AND, OR) were applied through the advanced search interfaces of IEEE Xplore, ACM Digital Library, ScienceDirect, and PubMed. Because the intersection between fNIRS and game-integrated learning is highly specific, we used five operational search queries: (1) ``game-based learning'' and fNIRS, (2) ``educational game'' and fNIRS, (3) ``serious game'' and fNIRS, (4) (``serious game'' or ``educational game'' or ``game-based learning'') and ``fNIRS'', and (5) gamification and fNIRS. The same five queries were used across the three search phases, and the complete search strategy and retrieved record counts are reported in Appendix 1,~\autoref{tab:supp_search_strategy}.

\begin{itemize}
    \item \textbf{Inclusion Criteria:} \begin{itemize}
        \item  Study used fNIRS as a primary measurement tool
 \item  Study was published in English
 \item  Study employed a game-integrated environment
 \item  Study was empirical in design (e.g., experiments, user studies)
 \item Study focused on pedagogical or initial skill acquisition (e.g., academic, medical training, or educational contexts)
 \item Study targeted cognitive or pedagogical learning outcomes (e.g., memory, attention, skill acquisition)
    \end{itemize}
     \item  \textbf{Exclusion Criteria:}
\begin{itemize}
 \item Studies with no fNIRS measurements or reported results
 \item Non-English publications
 \item Studies with no game-integrated learning environment
 \item Review papers, systematic reviews, or non-empirical studies
 \item Purely entertainment-focused games without explicit learning objectives
 \item Rehabilitation-focused systems where the primary objective was recovery rather than skill acquisition
 \item Studies that were outside the scope of educational or cognitive learning contexts, even if fNIRS was used
 \item Theses, dissertations, books, book chapters, preprints, and conference abstracts without a peer-reviewed full paper
\end{itemize}
\end{itemize}

To ensure conceptual consistency, studies that did not clearly align with both game-based learning and fNIRS research were excluded during later screening. Borderline cases such as \citet{brainsci11010106} and \citet{Lingelbach2023TowardsUserAware} were retained because they satisfied the inclusion criteria and reported empirical fNIRS findings in learning-related game environments. Although gamification is frequently discussed in the GILS literature, no empirical fNIRS studies explicitly framed as gamification-based learning systems met our inclusion criteria.

\subsection{Search Procedure}
The search was conducted in three phases: June 13–16, 2024; June 2, 2025; and April 1, 2026. The same search queries, eligibility criteria, and screening procedure were applied across all three phases. Records retrieved from the five sources were consolidated for screening. Potential duplicate records were identified by matching article titles and author information and were then manually cross-checked to confirm that they referred to the same publication before removal. The remaining records were screened according to the predefined inclusion and exclusion criteria. The numbers of records identified, screened, excluded, and included in each phase are reported in \autoref{sec:selection of sources of evidence} and summarized in the PRISMA-ScR flow diagram.

\subsection{Data Extraction and Verification}

Primary screening was conducted by two reviewers using a standardized checklist. Inter-rater agreement was assessed using Cohen's kappa ($\kappa > 0.70$); disagreements were resolved through discussion among the broader team of four reviewers. Ambiguous cases and the final included papers were verified by an additional author.

One researcher with experience in fNIRS and game-integrated learning conducted data extraction for all included studies. The extracted study-level information covered learning platform, multimedia delivery platform, outcome measure, study type, sample size and participant characteristics, study design and duration, ecological setting, fNIRS instrumentation and cortical coverage, preprocessing procedures, statistical approaches, and reporting of formal power analysis. Additional reviewers subsequently cross-checked the extracted information against the primary studies. This verification was used to identify and correct extraction errors; no substantive disagreements regarding device classification, outcome categorization, or study-type assignment were identified during this verification process.

Disagreements during study screening primarily concerned whether the game-based activities had an explicit learning or skill-acquisition purpose rather than being used in contexts where learning was not the primary objective. These cases were resolved through discussion using the predefined inclusion and exclusion criteria. A separate raw count of screening disagreements and a category-level breakdown of those disagreements were not tabulated during the review process; therefore, we do not retrospectively estimate these values.  No review protocol was prospectively registered for this scoping review.

\section{Results}\label{sec:results}

The review included 20 studies after the screening and eligibility
process described in~\autoref{sec:methodology} and summarized in
the PRISMA flow diagram. The included studies varied in learning
context, multimedia delivery platform, study design, outcome measures,
fNIRS instrumentation, cortical coverage, and analytical procedures.
The results are reported at two complementary levels below: a
corpus-level synthesis of patterns across the included studies and a
more detailed examination of individual studies.
The characteristics of the included studies are
reported in~\autoref{sec: Classification}, including learning
platforms, multimedia delivery platforms, outcome measures, study
designs, and fNIRS instrumentation. ~\autoref{sec:high level}
provides a corpus-level synthesis of patterns across the included
studies, while~\autoref{sec:detailed} provides a more detailed
study-level examination of selected methodological and
neurophysiological findings. 

\subsection{Selection of Sources of Evidence}\label{sec:selection of sources of evidence}
\begin{figure}[htb]
\centering
  \includegraphics[width=0.8\textwidth]{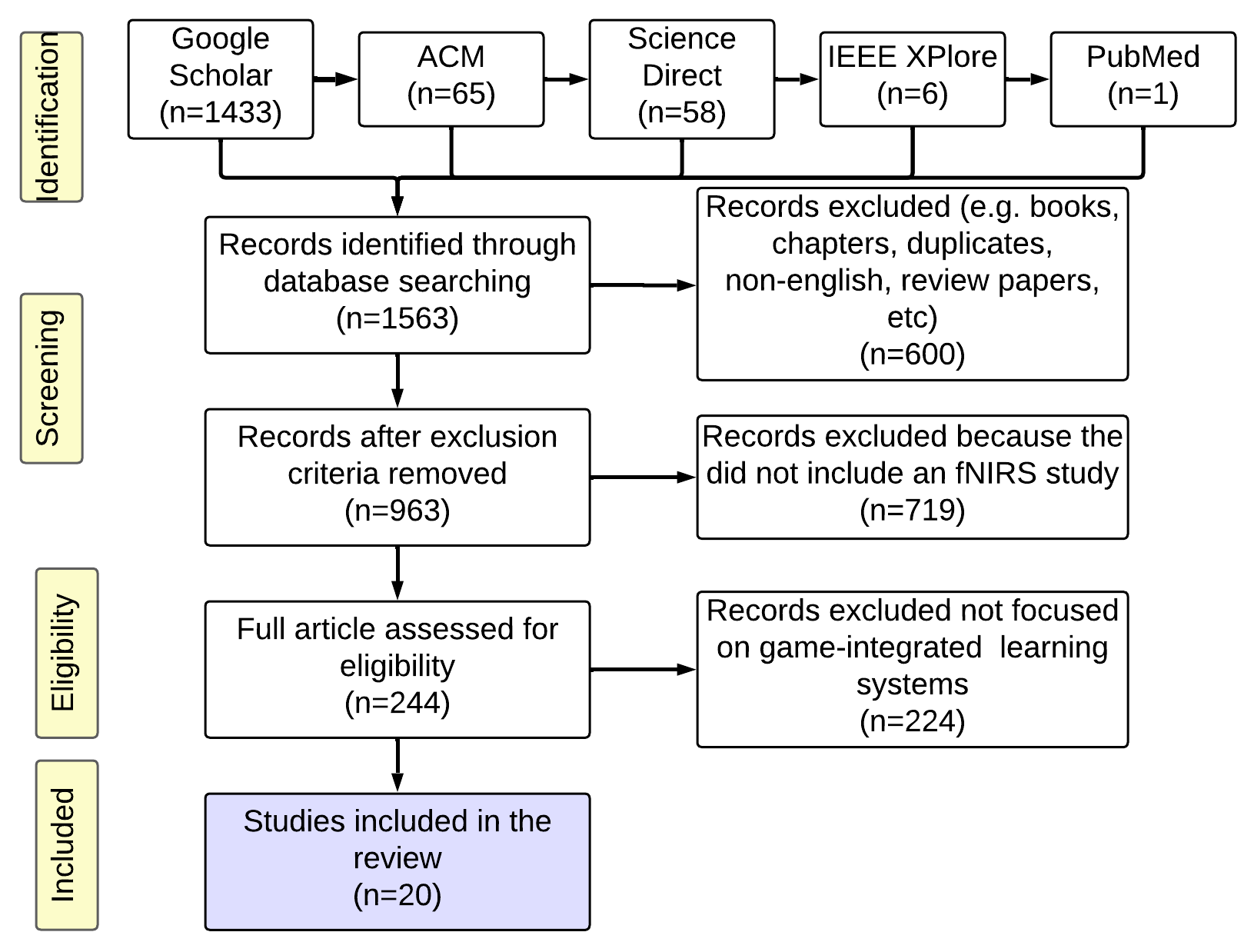}
  \caption{PRISMA-ScR flowchart summarizing the study identification, screening, eligibility assessment, and inclusion process based on the combined results of the initial search and subsequent search updates.}

  \label{fig:methodology}
\end{figure}

The complete study-selection process across the three search phases is summarized in the PRISMA-ScR flow diagram (\autoref{fig:methodology}). The initial search conducted from June 13–16, 2024 identified 957 records: IEEE Xplore ($n = 6$), PubMed ($n = 1$), ACM Digital Library ($n = 35$), ScienceDirect ($n = 27$), and Google Scholar ($n = 888$). After excluding books, chapters, duplicates, non-English publications, review papers, and other ineligible records $(n = 260)$, $697$ records remained. Of these, $510$ records were excluded because they did not include an fNIRS study, leaving $187$ full-text articles for high-level analysis. A further $170$ records were excluded because they were not focused on learning in game-integrated learning systems, resulting in $17$ included studies.
On June 2, 2025, an updated search was conducted using the same databases, the same keyword strategy, and eligibility criteria. This update identified additional records from ACM Digital Library $(n = 29)$, ScienceDirect $(n = 27)$, and Google Scholar $(n = 246)$, while no new records were identified from IEEE Xplore or PubMed at that stage. After excluding duplicates, books, chapters, non-English publications, review papers, and other non-eligible records $(n = 261)$, $41$ records remained. Of these, $14$ records were excluded because they did not include an fNIRS study, leaving $27$ full-text articles for high-level analysis. A further $25$ records did not meet the learning-context criterion, resulting in 2 newly included studies. This brought the cumulative total to $19$ studies.

 A final search update was conducted on April 1, 2026. This search identified newly added records from ACM Digital Library $(n = 1)$, ScienceDirect $(n = 4)$, and Google Scholar $(n = 299)$. After removing duplicates, chapters, review papers, and other ineligible records $(n = 79)$, $225$ records remained for screening. Of these, $195$ were excluded because they did not include an fNIRS study. The remaining $30$ records were assessed against the GILS criterion, of which $29$ were excluded, resulting in one newly included study \cite{sharmin2024fnirs}. This brought the final total to $20$ studies.

At the end of the search process, the cumulative records identified across all stages were IEEE Xplore $(n = 6)$, PubMed $(n = 1)$, ACM Digital Library $(n = 65)$, ScienceDirect $(n = 58)$, and Google Scholar $(n = 1433)$. Google Scholar was used as a supplementary search source to improve coverage; therefore, its records were not treated as mutually exclusive with those of IEEE Xplore, ACM Digital Library, ScienceDirect, or PubMed, and duplicate records were removed during screening.

\subsection{Classification Characteristics of Sources of Evidence}\label{sec: Classification}
To support a structured synthesis of the selected studies, we developed a classification framework that captures the main characteristics of fNIRS-enabled GILS. The classification dimensions were refined through repeated review of the included papers to reflect both methodological choices and design-relevant characteristics of GILS. Although individual studies may relate to more than one category, primary classifications were used where appropriate to support consistent synthesis. Device type and outcome measures could include multiple categories when a study used hybrid platforms or reported multiple outcome types. The framework comprises four dimensions: learning platform, device type, outcome measure, and study type, each corresponding to a column in \autoref{tab:listed paper}.
\begin{longtable}{p{2.5 cm}|p{0.6cm}|p{2.5cm}|p{1.5cm}|p{2cm}|p{1.7cm}}
\caption{A list of the selected 20 papers for in-depth reviews and their classifications. Publication years of the included studies (2014–2025), based on searches updated through April 2026. No included study was published in 2026.}
\label{tab:listed paper}\\

\hline
Title & Year & Learning Platform & Device Type & Outcome Measure & Study Type\\
\hline
\endfirsthead

\caption{Continued}\\
\hline
Title & Year & Learning Platform & Device Type & Outcome Measure & Study Type\\
\hline
\endhead

\hline
\endfoot

The potential use of neurophysiological signals for learning analytics \cite{Ninaus2014PotentialUse} 
& 2014 & Serious Game (Cognitive Assessment: ``Rule Learning, U get it U catch it'')& Screen-based (Computer) & Brain activation \& behavioral & Cognitive response \\ \hline

An optical brain imaging study on the improvements in mathematical fluency from game-based learning \cite{Cakir2015AnOptical} 
& 2015 & Game-based learning (Math: ``Arithmetic Fluency, MathDash'') & Screen-based (Tablet) & Brain activation, learning outcome, subjective measure\& behavioral & Comparative \\ \hline

Comparison of virtual reality and hands-on activities in science education via functional near-infrared spectroscopy \cite{LAMB201814} 
& 2018 & Educational Game (Biology: ``DNA Replication'') & Immersive \& Specialized (VR \& hands-on) & Brain activation, learning outcome, \& behavioral & Comparative \\ \hline

The brain's response to digital math apps: A pilot study examining children's cortical responses during touch-screen interactions \cite{BakeMoyeTuck2018ac} 
& 2018 & Educational Game (Math: ``Montessori Numbers'', ``Motion math'', ``LetsTans Kids'') & Screen-based (iPads) & Brain activation & Cognitive response \\ \hline

Performance monitoring via functional near-infrared spectroscopy for virtual reality-based basic life support training \cite{Aksoy2019PerformanceMonitoring} 
& 2019 & Serious Game (Medical Training: ``Basic Life Support'') & Immersive (VR) & Brain activation, knowledge gain, \& behavioral & Comparative \\ \hline

An implicit dialogue injection system for interruption management \cite{Shibata2019ImplicitDialogue} 
& 2019 & Game-based Learning (Cognitive Assessment: ``Memory Performance'') & Screen-based (Computer) & Brain activation, subjective measure, \& behavioral & Comparative \\ \hline

Assessing correlation between virtual reality-based serious gaming performance and cognitive workload changes via functional near-infrared spectroscopy \cite{Aksoy2019Correlation} 
& 2019 & Serious Game (Medical Training: ``Basic Life Support'') & Immersive (VR) & Brain activation, learning outcome, \& behavioral & Comparative \\ \hline

Assessing intravenous catheterization simulation training of nursing students using functional near-infrared spectroscopy \cite{Aksoy2020AssessingIntravenous} 
& 2020 & Serious Game (Medical Training: ``Intravenous Catheterization'')& Specialized (Virtual I.V. simulator) & Brain activation, learning outcome, \& behavioral & Comparative \\ \hline

The effects of two game interaction modes on cortical activation in subjects of different ages: A functional near-infrared spectroscopy study \cite{Ge2021EffectsTwogames} 
& 2021 & Game-based Learning (Cognitive Assessment: ``Obstacle Avoidance'') & Immersive (VR vs. Mobile; VR control group so considered as Immersive group) & Brain activation \& behavioral& Comparative \\ \hline

What can ``drag \& drop'' tell? Detecting mild cognitive impairment by hand motor function assessment under dual-task paradigm \cite{ZHANG2021102547} 
& 2021 & Serious Game (Cognitive Assessment: ``cogSYS'' & Specialized (NanoPi M4) & Brain activation, subjective measure, \& behavioral & Cognitive response \\ \hline

Single-trial recognition of video gamers' expertise from brain hemodynamic and facial emotion responses \cite{brainsci11010106} 
& 2021 & Game-based Learning (Chemistry: ``Valence \& Formula Learning'')& Screen-based (Game video) & Brain activation \& subjective measure & Cognitive response \\ \hline

Neural correlates of cognitive load while playing an emergency simulation game: A functional near-infrared spectroscopy (fNIRS) study \cite{Sevecenko2022NeuralCorrelates} 
& 2022 & Serious Game (Emergency Simulation) & Screen-based (Computer) & Brain activation, subjective measure, \& behavioral & Cognitive response \\ \hline

Emotional responses to performance feedback in an educational game during cooperation and competition with a robot: Evidence from fNIRS \cite{Lei2023EmotionalResponse} 
& 2023 & Educational Game (Cognitive Assessment: ``Memory Performance'')& Specialized (Humanoid robot) & Brain activation, subjective measure, \& behavioral & Comparative \\ \hline

Towards user-aware VR learning environments: Combining brain-computer interfaces with virtual reality for mental state decoding \cite{Lingelbach2023TowardsUserAware} 
& 2023 & Serious Game (Cognitive Assessment: ``Memory Performance'')& Immersive (VR) & Brain activation, subjective measure, \& behavioral & Comparative \\ \hline

Engaging learners with games—Insights from functional near-infrared spectroscopy \cite{de2023engaging} 
& 2023 & Game-based Learning (Math: ``fraction'') & Screen-based (Computer) & Brain activation, subjective measurement, \& behavioral & Comparative \\ \hline

Adaptive computerized cognitive training decreases mental workload during working memory precision task: A preliminary fNIRS study \cite{LANDOWSKA2024103206} 
& 2024 & Game-based learning (Cognitive Assessment: ``Memory Performance'') & Immersive (AR) & Brain activation, knowledge gain, subjective measure, \& behavioral & Comparative \\ \hline

Cognitive engagement for STEM+C education: Investigating serious game impact on graph structure learning with fNIRS \cite{Shayla2024StemC} 
& 2024 & Educational Game (Math: Graph Structure) & Screen-based (Computer) & Brain activation, knowledge gain, subjective measure, \& behavioral & Comparative \\ \hline


Dyads composed of members with high prior knowledge are most conducive to digital game-based collaborative learning \cite{GUI2025105266} 
& 2025 & Game-based Learning (Chemistry: ``Minecraft'') & Screen-based (Computer) & Brain activation, knowledge gain, subjective measure, \& behavioral & Comparative \\ \hline

Neural investigations of a digital game-based intervention for young learners with mathematical developmental variability \cite{WANG2025272} 
& 2025 & Educational Game (Math: Numerical Magnitude \& Addition) & Screen-based (Computer) & Brain activation, learning outcome, \& behavioral & Comparative \\ \hline

Functional near-infrared spectroscopy (fNIRS) analysis of interaction techniques in touchscreen-based educational gaming \cite{sharmin2024fnirs} 
& 2025 & Educational Game (Math: Graph Structure) & Screen-based (Computer) & Brain activation, learning outcome, subjective measure \& behavioral & Comparative \\

\end{longtable}
\subsubsection{Learning Platform} 
The literature on GILS employs diverse, sometimes interchangeable, terminology such as game-based learning, educational game, and serious game \cite{becker2021s}. To enable systematic comparison, we categorized the reviewed studies according to their primary pedagogical purpose rather than relying solely on author-assigned labels. Three categories were identified: game-based learning, educational games, and serious games.
\begin{itemize}
    \item [-] \textbf{Serious Game: } These games focus on training, simulation, or skill development in applied or high-stakes contexts, extending beyond entertainment to professional or clinical competencies. Representative examples include medical simulation systems such as the basic life support training platform \citep{Aksoy2019PerformanceMonitoring} and the intravenous catheterization simulator \citep{Aksoy2020AssessingIntravenous}.
    \item [-] \textbf{Game-based Learning: } This refers to the use of digital or non-digital games, originally designed for either entertainment or open-ended exploration, and used in education to support learning objectives through gameplay and problem-solving.  Example: Minecraft \cite{GUI2025105266} and MathDash \cite{Cakir2015AnOptical}.
    \item [-] \textbf{Educational Game: } These games are designed for structured educational content while integrating learning objectives directly into gameplay mechanics with enjoyment. These games align closely with curricula or targeted learning outcomes, such as the Montessori Numbers Game for Kids \cite{BakeMoyeTuck2018ac}.

 \end{itemize}

\subsubsection{Multimedia Delivery Platform (Device Type)}

To characterize the multimedia contexts in which learning activities and fNIRS measurements were conducted, multimedia delivery platforms were classified into three categories reflecting differences in interaction fidelity and immersion level. These categories are reflected in the ‘Device Type’ column of \autoref{tab:listed paper}. Most studies were assigned to a single device category; however, studies involving hybrid configurations could be represented in more than one category. 
\begin{itemize}
    \item [-] \textbf{Screen-based:} These types of displays encompass screen-based systems, including desktop computers, tablets, and iPads. This is the most frequently used device category across the reviewed literature. 
    \item [-] \textbf{Immersive:} These include VR and AR systems that offer higher levels of immersion and spatial interaction. VR platforms used in the reviewed studies included systems such as the 3DMedSim module and HTC Vive, Oculus, and AR platforms. 
    \item [-] \textbf{Specialized:} Other devices capture interaction modalities that do not conform to standard screen-based or head-mounted interfaces, including hands-on clinical simulators (e.g., the Virtual IV Simulator), humanoid robots, and specialized embedded platforms (e.g., the NanoPi M4). 

 \end{itemize}

\subsubsection {Outcome Measurement}
Outcome measures were grouped into four categories reflecting the main evaluation goals of the reviewed studies. Most studies employed multiple outcome categories concurrently. This reflects interest in triangulating neural, behavioral, and experiential dimensions of learning.
\begin{itemize}

    \item [-] \textbf{Brain Activation Analysis:} The selected studies primarily investigated how specific brain regions, particularly the prefrontal cortex, responded in GILS. The focus was on capturing changes in neural activation patterns (e.g., using oxygenated and deoxygenated hemoglobin) in relation to different learning tasks or game conditions. All 20 reviewed studies included brain activation as an outcome measure. These measures complement behavioral outcomes, supporting cautious interpretation of task-related mental effort and learning.

 
    \item[-] \textbf{Behavioral Performance Metrics:} These metrics capture in-game or task-based performance indicators such as in-game performance logs, response time, task completion rate, error rate, and accuracy. These were the most consistently reported outcomes alongside brain activation data.

   \item[-] \textbf{Subjective Measures:} These include self-reported instruments such as the NASA task load (NASA-TLX) \cite{hart1988development}, System Usability Scale (SUS) \cite{brooke2013sus}, and customized questions measuring engagement or affect ratings. Subjective measures were frequently combined with brain and behavioral data in comparative studies.
      \item [-] \textbf{Learning Outcome: } \ This outcome encompasses pre- and post-assessment of domain knowledge or skill, including quiz scores, knowledge tests, and task mastery assessments. Half of the reviewed studies included this category ($n=10$).

 \end{itemize}
 \subsubsection{Study Type}
The reviewed studies were classified into two primary types based on their dominant research focus. This distinction clarifies whether fNIRS was deployed primarily as an evaluation tool to compare outcomes across conditions, or as an exploratory mechanism to characterize cognitive and emotional states during interaction.
 \begin{itemize}
    \item [-] \textbf{Comparative studies:} This section includes articles that compare two groups, two materials, or two tasks to examine differences in outcomes.
    Comparative studies evaluate differences in outcomes across learning conditions, instructional approaches, interaction devices, or learner groups. For example, comparison of game-based and non-game-based learning contexts \cite{de2023engaging, Shayla2024StemC}, or comparing input devices while playing an educational game \cite{sharmin2024fnirs}.
    \item [-] \textbf{Cognitive response studies:} 
Cognitive response studies examined task-related cortical hemodynamic responses in relation to cognitive, emotional, workload-related, or expertise-related constructs, where any experimental conditions served to induce varying cognitive states rather than to compare outcomes across them \cite{Ninaus2014PotentialUse, Sevecenko2022NeuralCorrelates}. 
 \end{itemize}

\subsection{Corpus-Level Synthesis}\label{sec:high level}
This section presents a high-level quantitative analysis of the 20 selected studies, focusing on publication patterns, learning platforms, interaction modalities, study types, and outcome measures. The analysis is intended to characterize the current research landscape in fNIRS-enabled game-integrated learning. 

\subsubsection{Publication Timeline and Learning Platforms}

\begin{figure}[h]
    \centering
    \includegraphics[width=\linewidth]{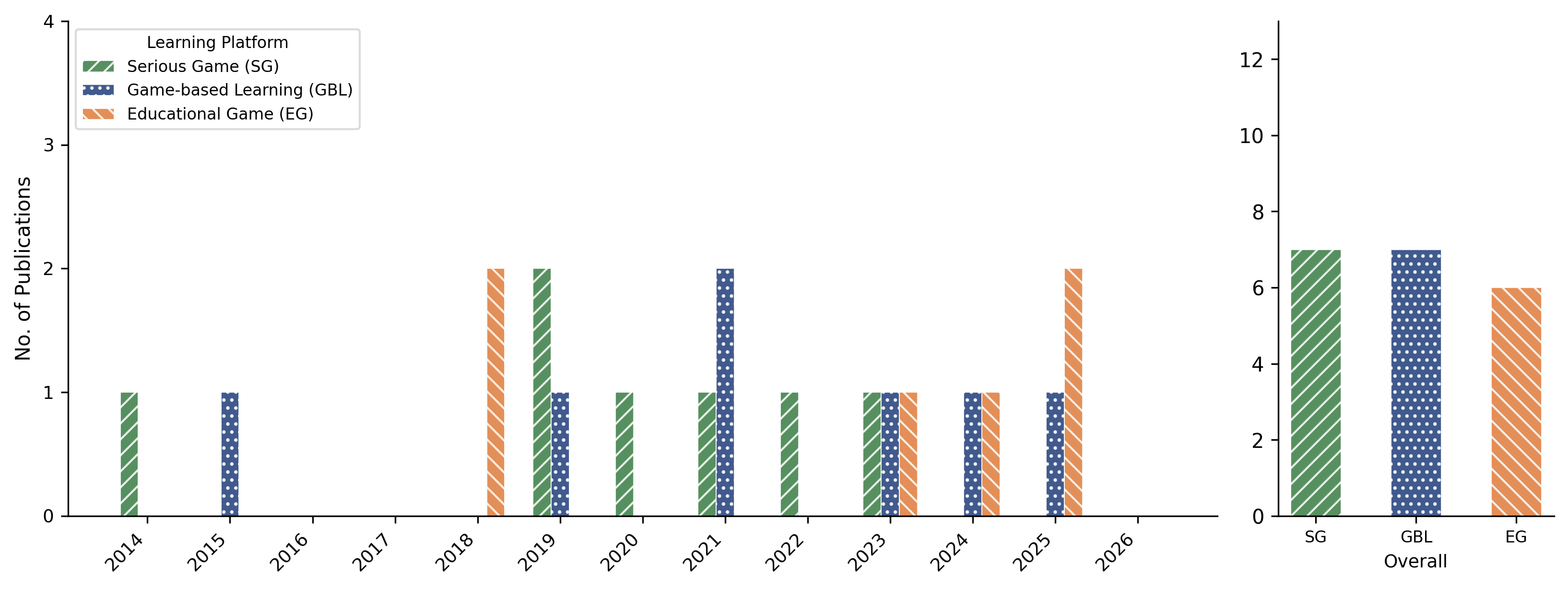}
    \caption{Publications per year by game-integrated learning system (GILS). The final search was updated through April 1, 2026; no included study was published in 2026.}
    \label{fig:fig3_platform}
\end{figure}

Although the search was updated through April 2026, the 20 included studies were published between 2014 and 2025. Nine of the 20 included studies were published from 2022 onwards (\autoref{fig:fig3_platform}).
Across the review period, serious games ($n = 7$) and game-based learning studies ($n = 7$) were equally common. Serious games were used primarily in professional training contexts, whereas game-based learning studies were applied across a broader range of open-ended learning environments. Educational games followed ($n = 6$), typically in structured subjects such as mathematics and science.
The substantial use of serious games may reflect the methodological advantages of purpose-built simulation environments, which provide standardized task structures, defined performance metrics, and greater experimental control. These characteristics can facilitate clearer interpretation of fNIRS signals. Game-based learning environments often involve more open-ended interactions and variable learner behaviors, which may introduce additional challenges for isolating learning-related neural activation.

These categories were also associated with distinct educational applications. Serious games were used for professional skill training, where the connection between gameplay and real-world performance was explicit. Educational games were concentrated in mathematics and other structured subject areas, whereas game-based learning studies were distributed across a wider range of educational contexts. Together, these patterns illustrate the versatility of GILS, while also highlighting the challenge of synthesizing findings across studies that target different learning objectives and learner populations.

Notably, a small but meaningful subset of studies extended GILS research to learners with specific cognitive or developmental needs, including mild cognitive impairment \cite{ZHANG2021102547} and mathematical developmental variability \cite{WANG2025272}. Although limited in number, these studies point toward promising opportunities for more accessible and inclusive game-integrated learning designs.

\subsubsection{Multimedia Delivery Platform (Device Type) and Interaction Modalities}

\begin{figure}[h]
    \centering
    \includegraphics[width=\linewidth]{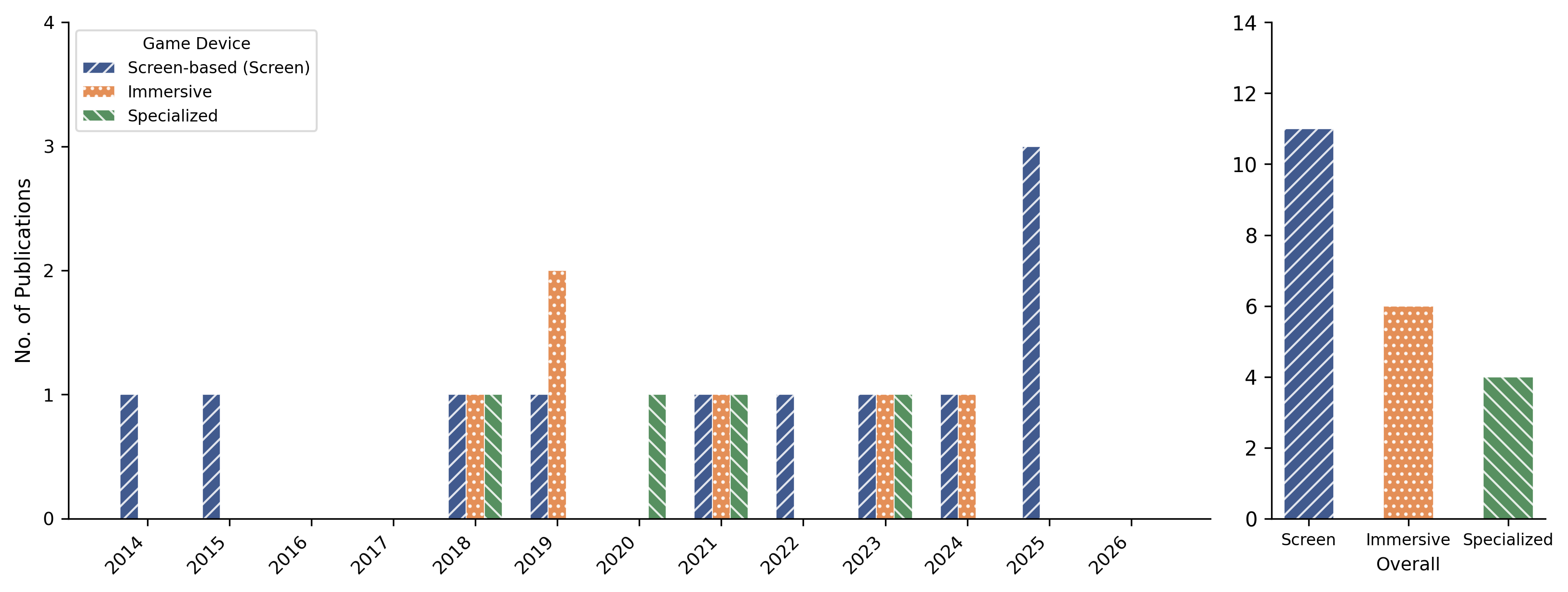}
    \caption{Publications per year by device type. Search updated through April 1, 2026. Device categories are not mutually exclusive because one hybrid study [9] was classified as both immersive and specialized.}
    \label{fig:fig4_device}
\end{figure}

Screen-based devices ($n = 11$), including desktop computers, tablets, and iPads, dominate the reviewed literature across the review period (\autoref{fig:fig4_device}). Immersive platforms were represented in six studies, while specialized devices were represented in four studies. Because one study~\citet{LAMB201814} used a hybrid immersive-and-specialized configuration, these device-category counts are not mutually exclusive.
Specialized devices, such as simulators and robots, appear only intermittently and are typically confined to individual studies, suggesting that they represent emerging rather than established research directions. Their inclusion illustrates the range of interaction modalities examined in this literature, including embodied and multisensory learning contexts. However, the limited evidence does not establish a temporal shift toward these modalities or platform-specific effects on cortical hemodynamic responses.


Across the corpus, multimedia delivery platform and interaction modality
were closely related but were not interchangeable characteristics.
Screen-based studies generally involved conventional computer or
touch-based interaction, whereas immersive and specialized systems
incorporated more spatial, embodied, or task-specific forms of
interaction. Thus, the reviewed literature represents a range of
interaction contexts rather than a single form of game-integrated
learning. Because these studies also differed in learning task,
participant population, and fNIRS configuration, the distribution of
platforms and interaction modalities is interpreted descriptively rather
than as evidence that a particular platform produced a specific
neurophysiological response.

Outcome reporting also varied descriptively across device categories. Learning outcomes were reported in five of the 11 screen-based studies and four of the six immersive studies, while two of the four studies involving specialized devices reported learning outcomes. Behavioral measures accompanied fNIRS in nine of the 11 screen-based studies and in all studies involving immersive or specialized devices. Because one study used a hybrid immersive-and-specialized configuration, these device-category counts are not mutually exclusive. These distributions are descriptive and should not be interpreted as evidence that device type determined the choice or value of a particular outcome measure.

\subsubsection{Outcome Measures in Different Studies}

\begin{figure}[h]
    \centering
    \includegraphics[width=\linewidth]{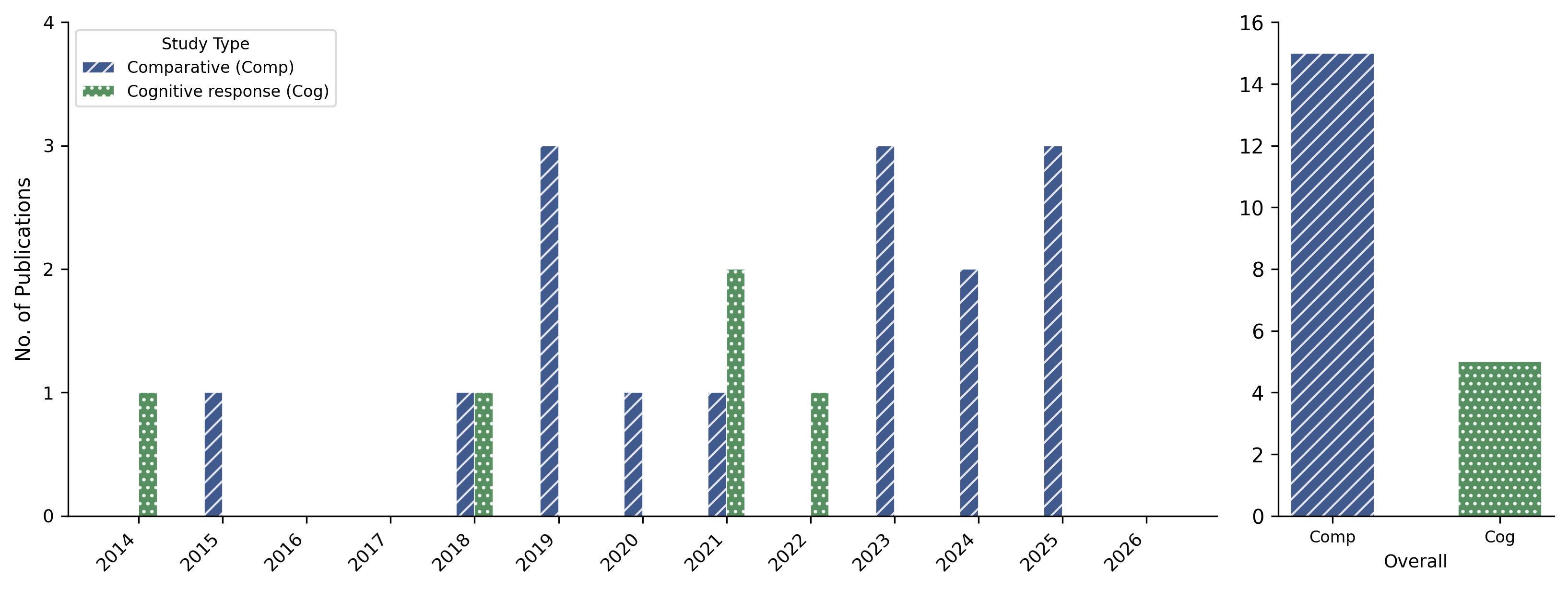}
    \caption{Publications per year by outcome measure. Search updated through April 1, 2026.}
    \label{fig:fig5_studytype}
\end{figure}

Regarding outcome measures (\autoref{fig:fig5_studytype}), all $20$ studies used brain activation analysis. 
Most studies also included behavioral performance measures ($n=18$), meaning that brain data were often considered alongside task performance.
About half of the studies ($n=10$) measured learning outcomes, such as knowledge gain or pre- and post-test results. 
Subjective measures, such as perceived workload and user experience, were used in more than half of the studies ($n=12$), mainly in studies focusing on cognitive load or user experience.

Learning outcomes were reported in 10 of the 20 included studies.
The remaining 10 studies primarily examined neurophysiological,
behavioral, or task-performance outcomes without reporting a separate
measure of learning. The primary studies generally did not state why
a learning outcome was omitted; therefore, we cannot determine from
the available reports whether this reflected study design priorities,
resource constraints, or broader field practices.


Taken together, the outcome measures show that the literature is weighted
toward neurophysiological and behavioral characterization, while direct
learning outcomes were assessed in only half of the included studies.
The frequent inclusion of behavioral measures indicates that fNIRS data
were commonly examined alongside observable task performance, whereas
learning and subjective outcomes were incorporated less consistently.
Consequently, the corpus provides more evidence about task-related
hemodynamic responses and concurrent performance than about whether
those responses correspond to changes in learning.

Outcome selection also differed by study type. All 10 studies that reported a learning outcome were classified as comparative studies, whereas none of the five cognitive response studies included a separate learning-outcome measure. Behavioral measures were reported in all 15 comparative studies and in three of the five cognitive response studies. Thus, comparative studies more often combined fNIRS with behavioral and learning measures, whereas the cognitive response studies primarily emphasized neurophysiological characterization of task- or state-related responses.
\subsubsection{Applied Study Types}

\begin{figure}[h]
    \centering
    \includegraphics[width=\linewidth]{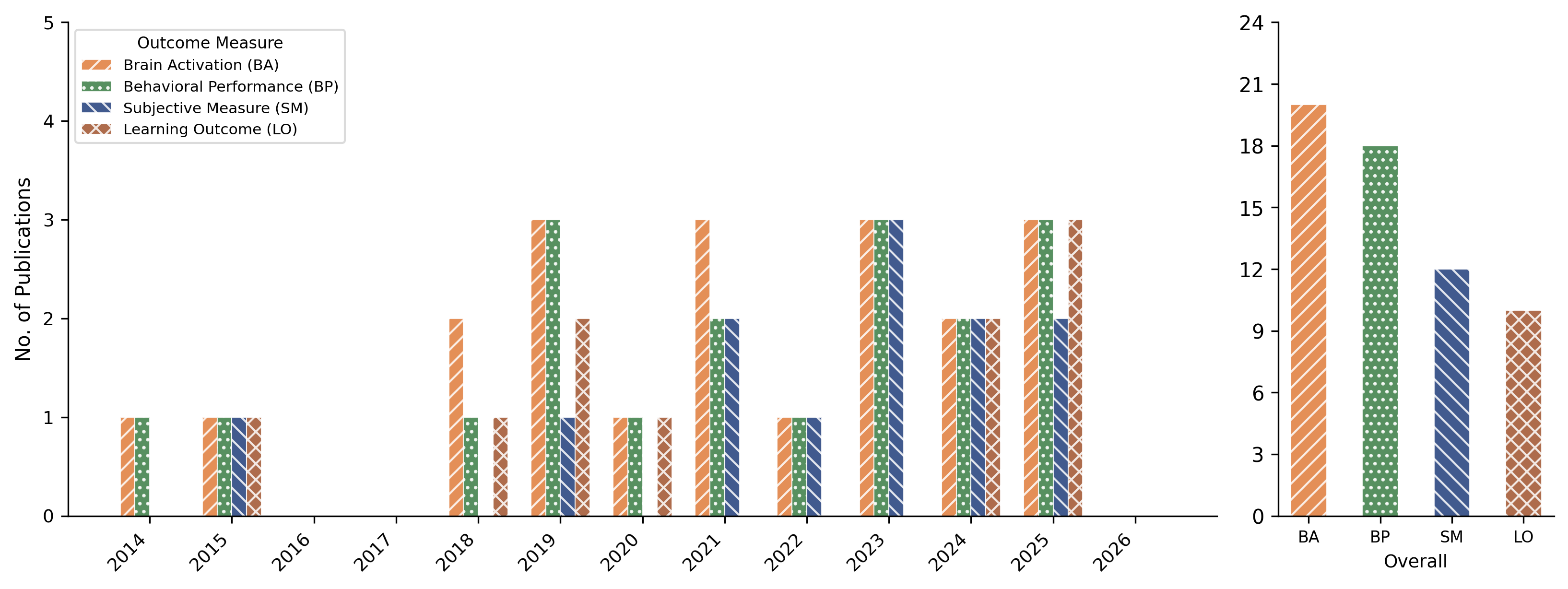}
    \caption{Publications per year by fNIRS study type. Search updated through April 1, 2026.}
    \label{fig:fig6_outcomes}
\end{figure}

Comparative studies constitute the majority of the reviewed literature ($n=15$), with cognitive response studies representing a smaller but distinct subset ($n=5$) (\autoref{fig:fig6_outcomes}). Most comparative studies examined differences in task-related cortical hemodynamic responses across instructional conditions, interaction modes, task demands, or participant groups. A smaller set characterized hemodynamic responses in contexts where the primary studies related these responses to constructs such as engagement, workload, or expertise. 
Overall, the corpus therefore emphasizes comparisons between tasks,
conditions, or participant groups more than descriptive characterization
of neurophysiological responses alone. This distribution also means that
the reviewed studies address different analytical purposes: comparative
studies examine differences between predefined conditions or groups,
whereas cognitive response studies primarily characterize
responses occurring during particular learning or interaction contexts.
These study types are therefore complementary rather than directly
comparable categories of evidence.
The comparative studies themselves were heterogeneous in purpose, examining contrasts involving instructional conditions, interaction devices, training or experience levels, task demands, and learner groups. Therefore, the predominance of comparative designs should not be interpreted as representing a single common intervention contrast; rather, comparison was used across several distinct research questions within the corpus.

\subsubsection{fNIRS Devices, Configuration, and Analysis Techniques}

~\autoref{tab:study_design_analysis} summarizes the methodological configurations of the reviewed studies.

\begin{table}[h]
\centering
\caption{Summary of study designs and fNIRS analysis characteristics
across the included studies.}
\label{tab:study_design_analysis}

\begin{tabular}{p{3.0cm}p{8.2cm}}
\hline
\textbf{Characteristic} & \textbf{Distribution across studies} \\
\hline

Design &
Within-subject (8), between-subject (7), mixed (5) \\

Duration &
Cross-sectional (14), longitudinal (6) \\

fNIRS signal &
$\Delta$HbO only (8), $\Delta$HbO--$\Delta$HbR (11), OXY (1) \\

Statistical analyses &
Included $t$-tests, ANOVA, general linear models, non-parametric
tests, mixed-effects models, and other study-specific analyses \\

\hline
\end{tabular}
\end{table}
Within-subject designs were used in eight studies, between-subject designs in seven studies, and mixed designs in five studies. The mixed designs combined between-participant comparisons with repeated or within-participant factors, reflecting the range of experimental structures used across the reviewed corpus.
Most studies were cross-sectional ($n=14$), limiting inferences about how task-related cortical hemodynamic responses change over sustained engagement with game-integrated systems. 
The smaller subset of longitudinal studies ($n=6$) allowed
task-related hemodynamic responses to be examined across repeated
training or assessment sessions.

Across these longitudinal studies, repeated measurements ranged from
sessions separated by seven days to multi-week training protocols,
with some studies also including post-training or follow-up assessments.
Several studies reported changes in task-related cortical hemodynamic
responses across repeated sessions, sometimes alongside improvements in
behavioral or task-performance measures, whereas corresponding behavioral
changes were not consistently observed across all studies.
Together, these studies illustrate the use of fNIRS for tracking
within-participant neurophysiological changes across repeated training
or task exposure, although differences in session timing, populations,
and outcome measures limit a common interpretation of these changes.

With respect to fNIRS signals, combined oxygenated hemoglobin ($\Delta HbO$) – deoxygenated hemoglobin ($\Delta HbR$) analysis was the most prevalent approach ($n=11$). $\Delta HbO$ alone was analyzed in $8$ studies, and OXY-derived metrics by subtracting  $\Delta HbO$ and $\Delta HbR$ in $1$ study. Statistical approaches varied considerably, ranging from parametric tests (t-tests, ANOVA, General Linear Model(GLM)) to mixed-effects models and non-parametric alternatives (Wilcoxon signed-rank, Mann–Whitney). This methodological diversity reflects the absence of standardized analysis pipelines, thereby reducing the comparability of findings across studies.

Eight commercial fNIRS system categories were represented across the reviewed studies, as summarized in~\autoref{tab:fnirs device}. The fNIRS Imager by Biopac was the most frequently represented system, appearing in eight studies. 
Systems varied in channel count, source-detector configuration, and analytical software. This variability reflects how researchers adapted their instrumentation to the specific demands of their learning contexts.

\begin{longtable}[h]{p{2.5cm}p{3.6cm}p{3.6cm}p{2cm}}
\caption{Specifications of fNIRS devices used in the included studies. 
ch denotes measurement channels, S sources, and D detectors. 
Where multiple studies used the same device family, reported configurations may vary across studies.}
\label{tab:fnirs device}
\\
\hline
Device & Manufacturer & Configuration & Software \\
\hline
\endfirsthead

\caption{Continued}\\
\hline
Device & Manufacturer & Configuration & Software \\
\hline
\endhead

\hline
\endfoot

fNIRS imager 1200/2000S
\cite{
LAMB201814,
Aksoy2019PerformanceMonitoring,
Cakir2015AnOptical,
brainsci11010106,
Aksoy2020AssessingIntravenous,
Aksoy2019Correlation,
Shayla2024StemC,
sharmin2024fnirs}
&
Biopac (USA)
&
Reported configurations varied across studies; 16 measurement locations were reported, while source--detector counts were not reported consistently across all studies
&
fNIRSoft \& COBI
\\

\hline

NIRSport2
\cite{
Sevecenko2022NeuralCorrelates,
Lingelbach2023TowardsUserAware,
de2023engaging,
WANG2025272}
&
NIRx Medical Technologies LLC (USA)
&
20--22 measurement ch (8S, 8D) in some studies;
15S and 14D, including 8 short-separation channels, in Lingelbach et al.; total measurement-channel count not explicitly reported
&
Aurora fNIRS;
MNE-Python, MNE-NIRS
\\

\hline

NIRSmart
\cite{Lei2023EmotionalResponse,Ge2021EffectsTwogames}
&
Danyang Huichuang Medical Equipment Co., Ltd. (China)
&
19 measurement ch (7S, 7D);
32 measurement ch (16S, 12D);
40 measurement ch (24S, 16D)
&
NirSpark
\\

\hline

ETG-4000
\cite{Ninaus2014PotentialUse,BakeMoyeTuck2018ac}
&
Hitachi Medical Co. (Japan)
&

44 measurement ch (16S, 14D; 22 channels per cap);
 24 measurement ch (10S, 8D; ETG-4000 52-channel system)
&
Hitachi Medical Co.
\\

\hline

Brite23 wireless fNIRS System
\cite{LANDOWSKA2024103206}
&
Artinis Medical Systems (Netherlands)
&
23 measurement ch (11S, 7D)
&
NIRS toolbox
\\

\hline

NirScan
\cite{ZHANG2021102547}
&
Danyang Huichuang Medical Equipment Co. Ltd. (China)
&
32 measurement ch (14S, 14D)
&
--
\\

\hline

Imagent
\cite{Shibata2019ImplicitDialogue}
&
ISS Inc. (USA)
&
16 measurement ch (8S, 2D)
&
Boxy
\\

\hline
NIRScout \cite{GUI2025105266}
&
NIRx Medical Technologies (USA)
&
20 measurement ch
&
nirsLAB \& MATLAB
\\

\hline

\hline

\end{longtable}

\autoref{tab:participant_analysis_characteristics}
summarizes participant characteristics, cortical
regions examined, preprocessing procedures,
and statistical approaches across the included studies.
Reporting of formal sample-size planning was limited across the corpus.
Four of the 20 included studies explicitly reported a formal power
analysis \cite{LAMB201814, Lei2023EmotionalResponse, WANG2025272,
sharmin2024fnirs}. For the remaining studies, a formal power analysis
was not identified in the primary reports, although some described
their work as exploratory or pilot research or provided other contextual
information about sample recruitment. The absence of a reported power
analysis should not be interpreted as evidence that no sample-size
planning was undertaken. These characteristics are considered
descriptively and were not used to classify individual studies as
methodologically strong or weak.

\begin{longtable}{p{1.45cm}p{0.85cm}p{2.0cm}p{2.0cm}p{2.55cm}p{2.05cm}}
\caption{Participant, neuroanatomical, preprocessing, and statistical characteristics of the included studies. \textbf{NR} indicates information not reported in the primary study.}
\label{tab:participant_analysis_characteristics}\\
\toprule
\textbf{Study} &
\textbf{N} &
\textbf{Population} &
\textbf{Brain region(s)} &
\textbf{Preprocessing} &
\textbf{Statistical approach} \\
\midrule
\endfirsthead

\multicolumn{6}{c}{\tablename\ \thetable\ -- continued}\\
\toprule
\textbf{Study} &
\textbf{N} &
\textbf{Population} &
\textbf{Brain region(s)} &
\textbf{Preprocessing} &
\textbf{Statistical approach} \\
\midrule
\endhead

\bottomrule
\endfoot

\citet{Ninaus2014PotentialUse} &
4 &
Adults, 21--31 years &
Left fronto-parietal network; central motor areas &
0.01--0.9 Hz filtering; motion-artifact correction (method not specified); 5-s pre-trial baseline &
Repeated-measures analysis of variance for behavioral data; descriptive/visual functional near-infrared spectroscopy analysis \\

\citet{Cakir2015AnOptical} &
27 &
Adults, 18--36 years; undergraduate/graduate students &
Prefrontal cortex (Brodmann areas 9, 10, 44, 45) &
0.1-Hz low-pass finite impulse response filter; sliding-window motion-artifact rejection; no explicit within-trial baseline correction reported &
Two-way mixed analysis of variance \\

\citet{LAMB201814} &
100 &
College-aged students; non-science majors &
Bilateral prefrontal cortex &
0.14-Hz low-pass; motion correction NR; pre/post exposure baselines; Z-score standardization &
Repeated-measures analysis of variance; one-way analysis of variance; Tukey honestly significant difference test; Pearson correlation; $t$-test \\

\citet{BakeMoyeTuck2018ac} &
10 &
Second-grade children, 7.4--8.5 years &
Bilateral intraparietal sulcus; right angular gyrus; bilateral dorsolateral prefrontal cortex &
0.5-Hz high-pass and 0.01-Hz low-pass filtering
(as reported in the primary study); motion-artifact
handling NR; baseline correction NR; oxygenated hemoglobin analyzed
using general-linear-model-derived standardized beta coefficients &
One-sample $t$-tests; no multiple-comparison correction reported \\

\citet{Aksoy2019PerformanceMonitoring} &
22 &
Healthcare and non-healthcare participants; age NR &
Prefrontal cortex; left prefrontal cortex channel 3 as region of interest &
0.1--0.15-Hz low-pass finite impulse response filter; sliding-window motion-artifact rejection; 10-s pre-task baseline &
Kruskal--Wallis; Mann--Whitney U; Shapiro--Wilk \\

\citet{Shibata2019ImplicitDialogue} &
10 &
Adults, 18--35 years &
Bilateral prefrontal cortex &
0.5-Hz low-pass; 30-s moving-window baseline; 3-min session baseline &
Aligned Rank Transform analysis of variance; Wilcoxon signed-rank; support vector machine classification \\

\citet{Aksoy2019Correlation} &
11 &
Healthcare and non-healthcare adults; age NR &
Prefrontal cortex; channel 3 as region of interest &
Low-pass finite-impulse Hamming filter (cutoff NR); sliding-window motion-artifact rejection; 10-s baseline &
Wilcoxon signed-rank; Mann--Whitney U \\

\citet{Aksoy2020AssessingIntravenous} &
23 &
Novice nursing students and senior nurses &
Left prefrontal cortex &
Low-pass finite-impulse Hamming filter (cutoff NR); sliding-window motion-artifact rejection; baseline NR &
Wilcoxon signed-rank; Mann--Whitney U; Shapiro--Wilk \\

\citet{Ge2021EffectsTwogames} &
31 &
Young and older adults &
Left/right prefrontal cortex; left/right motor cortex &
Butterworth band-pass (range NR); spline interpolation for motion artifacts; hemodynamic-response-function baseline $-2$ to 0 s &
Shapiro--Wilk; paired $t$-tests; Pearson correlation \\

\citet{ZHANG2021102547} &
37 &
Older adults; mild cognitive impairment and healthy-control groups &
Left/right prefrontal cortex; left/right motor cortex; occipital lobe &
Sixth-order Butterworth band-pass (range NR); motion artifacts removed (method NR); 3-min resting baseline &
Descriptive comparison of changes in oxygenated hemoglobin concentration; no formal inferential functional near-infrared spectroscopy statistics reported \\

\citet{brainsci11010106} &
30 &
Adults; novice, intermediate, and expert groups &
Bilateral prefrontal cortex &
0.14-Hz low-pass; channel-quality exclusion and $>3$ standard-deviation outlier removal; optical-density comparison with baseline &
Kruskal--Wallis with Dunn--Šidák correction; classification metrics with 10-fold cross-validation \\

\citet{Sevecenko2022NeuralCorrelates} &
27 &
Adults, 20--49 years &
Bilateral dorsolateral prefrontal cortex; left inferior frontal gyrus &
0.01--0.1 Hz band-pass; temporal derivative distribution repair; correlation-based signal improvement; Z-standardization; principal-component-analysis-based global-signal correction &
Linear mixed-effects models; likelihood-ratio tests \\

\citet{Lei2023EmotionalResponse} &
29 &
University students, 19--27 years &
Bilateral orbitofrontal cortex and rostral medial prefrontal cortex; right ventrolateral prefrontal cortex &
0.01--0.2 Hz band-pass; automatic motion correction; 30-s pre-task baseline &
Two-way repeated-measures analysis of variance; $t$-tests; Pearson correlation \\

\citet{Lingelbach2023TowardsUserAware} &
11 (9 completed all sessions) &
Adults, 24--34 years &
Frontopolar cortex; bilateral dorsolateral prefrontal cortex; premotor cortex; Broca area; motor cortex &
0.05--0.7 Hz band-pass; temporal derivative distribution repair; scalp-coupling-index rejection; short-channel general linear model regressors &
Linear mixed model with Bonferroni correction; area under the receiver operating characteristic curve/linear discriminant analysis; bootstrap confidence intervals \\

\citet{de2023engaging} &
41 &
Young adults, 18--30 years &
Middle/superior frontal and orbitofrontal regions &
0.01--0.50 Hz band-pass; channel-wise artifact detection; spline + wavelet correction; short-distance regression; 5-s pre-block baseline &
Mixed-effects models; paired $t$-tests; Holm correction; Cohen's $d$ \\

\citet{LANDOWSKA2024103206} &
85 &
Adults aged 45+ &
Bilateral dorsolateral prefrontal cortex &
Temporal derivative distribution repair; short-separation nuisance regressor; modified Beer--Lambert conversion; autoregressive iteratively reweighted least-squares beta estimation; filter range NR &
Mixed-effects model; generalized linear mixed model; false discovery rate correction; contrasts; Wilcoxon signed-rank \\

\citet{Shayla2024StemC} &
12 &
Graduate students; adults $>18$ years &
Lateral prefrontal cortex and ventromedial prefrontal cortex &
Filter NR; motion correction NR; baseline NR &
Independent-samples $t$-test; one-way analysis of variance; descriptive functional near-infrared spectroscopy statistics \\

\citet{GUI2025105266} &
106 &
College students; non-chemistry majors &
Left prefrontal cortex; right temporoparietal junction &
0.01--0.7 Hz task band; physiological-frequency exclusion; no additional motion correction; first/last 10 s excluded &
Independent-samples $t$-test with false discovery rate correction; Pearson correlation; bootstrap mediation; analysis of variance; multivariate analysis of variance \\

\citet{WANG2025272} &
32 &
First-grade children; dyscalculia-at-risk and non-risk groups &
Bilateral inferior frontal gyrus, intraparietal sulcus, and superior frontal gyrus &
Filter NR; motion correction NR; baseline NR &
Repeated-measures general linear model; two-way analysis of variance; simple effects; independent-samples $t$-test \\

\citet{sharmin2024fnirs} &
16 &
Graduate students, 23--32 years &
Bilateral dorsolateral and anteromedial prefrontal cortex &
Modified Beer--Lambert law conversion; filter range NR; no computational motion-artifact correction reported; 20-s initial rest &
General linear model coefficient test; Mann--Whitney U; Shapiro--Wilk; paired $t$-tests \\

\end{longtable}

To provide study-level methodological information relevant to bias and interpretability,~\autoref{tab:design_hardware_characteristics} summarizes the sample size, study design and duration, ecological setting, and whether a formal power analysis was reported for each included study.

\begin{longtable}{p{1.65cm}p{0.85cm}p{3.10cm}p{2.85cm}p{2.15cm}}

\caption{Study-level methodological characteristics of the included studies, including sample size, study design and duration, ecological setting, and reporting of formal power analysis. ``Not identified'' indicates that a formal power analysis was not identified in the primary report and should not be interpreted as evidence that no sample-size planning was undertaken.}
\label{tab:design_hardware_characteristics}\\

\toprule
\textbf{Study} &
\textbf{N} &
\textbf{Design and duration} &
\textbf{Ecological setting} &
\textbf{Formal power analysis} \\
\midrule
\endfirsthead

\multicolumn{5}{c}{\tablename\ \thetable\ -- continued}\\
\toprule
\textbf{Study} &
\textbf{N} &
\textbf{Design and duration} &
\textbf{Ecological setting} &
\textbf{Formal power analysis} \\
\midrule
\endhead

\bottomrule
\endfoot

\citet{Ninaus2014PotentialUse} &
4 &
Within-subject; single session ($\sim$30 min) &
Laboratory; simple 2D game &
Not identified \\

\citet{Cakir2015AnOptical} &
27 &
Between-subject; single pre/game-or-control/post session &
University laboratory &
Not identified \\

\citet{LAMB201814} &
100 &
Between-subject; single session &
Controlled university setting &
Yes \\

\citet{BakeMoyeTuck2018ac} &
10 &
Within-subject; single session &
University research center; iPad interaction &
Not identified \\

\citet{Aksoy2019PerformanceMonitoring} &
22 &
Between-subject; longitudinal, Day 1 and Day 7 &
Controlled VR/simulation laboratory &
Not identified \\

\citet{Shibata2019ImplicitDialogue} &
10 &
Within-subject; single session ($\sim$90 min) &
Laboratory &
Not identified \\

\citet{Aksoy2019Correlation} &
11 &
Mixed; longitudinal, Day 1 and Day 7 &
Laboratory/simulation center &
Not identified \\

\citet{Aksoy2020AssessingIntravenous} &
23 &
Mixed; longitudinal, Days 1, 2, 7, and 30 &
Laboratory/simulation center &
Not identified \\

~\citet{Ge2021EffectsTwogames} &
31 &
Mixed; single session with mobile and VR phases &
Design research laboratory &
Not identified \\

\citet{ZHANG2021102547} &
37  &
Between-subject; single session &
Laboratory &
Not identified \\

\citet{brainsci11010106} &
30 &
Mixed; between-subject expertise groups with repeated within-participant trials; single session &
Laboratory; passive gameplay-video observation &
Not identified \\

\citet{Sevecenko2022NeuralCorrelates} &
27 &
Within-subject; single session ($\sim$1 h) &
Quiet laboratory; emergency-simulation game &
Not identified \\

\citet{Lei2023EmotionalResponse} &
29 &
Within-subject; single session ($\sim$70--80 min) &
Controlled laboratory &
Yes \\

\citet{Lingelbach2023TowardsUserAware} &
11 (9 completed all sessions) &
Within-subject; longitudinal, 3 sessions ($\sim$90 min each) &
VR industrial simulation; not a real workplace &
Not identified \\

\citet{de2023engaging} &
41 &
Within-subject; single session ($\sim$1.5 h) &
Laboratory &
Not identified \\

\citet{LANDOWSKA2024103206} &
85  &
Between-subject; longitudinal, 8-week training + 5-week follow-up &
University Mixed Reality Lab; AR shopping transfer task &
Not identified \\

\citet{Shayla2024StemC} &
12 &
Between-subject; single session &
Controlled laboratory &
Not identified \\

\citet{GUI2025105266} &
106 &
Between-subject; single session (10-min learning phase) &
Controlled laboratory &
Not identified \\

\citet{WANG2025272} &
32 &
Mixed; longitudinal, 6 weeks with four weekly intervention sessions &
School setting with structured fNIRS protocol &
Yes \\

\citet{sharmin2024fnirs} &
16 &
Within-subject; single session with two counterbalanced quiz conditions &
University laboratory; touchscreen laptop &
Yes \\

\end{longtable}


Beyond device configuration, the included studies varied in their
preprocessing and approaches to signal-quality control
(\autoref{tab:participant_analysis_characteristics}). Reported approaches included
frequency filtering, baseline correction, motion-artifact correction,
and, in some studies, short-separation measurements or regression of
short-channel signals. These procedures were not reported consistently
across the corpus. We therefore distinguish between methods explicitly
reported by the primary studies and those for which sufficient
methodological information was not available, rather than assuming
equivalent control of motion, superficial hemodynamics, or systemic
physiological signals across studies.


Overall, the methodological landscape is heterogeneous at several levels:
studies differed in experimental design and duration, fNIRS system, and
channel configuration, cortical coverage, preprocessing procedures,
hemodynamic signals analyzed, and statistical approaches. Some
methodological choices were shared across subsets of studies, but no
single acquisition, preprocessing, or analysis pipeline characterized
the corpus as a whole. This variability is important when interpreting
findings across studies because differences in reported
neurophysiological responses may reflect not only the learning context
but also differences in measurement and analytical procedures.

As a descriptive sensitivity check, the three studies that explicitly used short-channel or short-separation regression procedures \citep{Lingelbach2023TowardsUserAware,de2023engaging, LANDOWSKA2024103206} each reported task- or condition-related prefrontal hemodynamic differences. This pattern indicates that prefrontal differences were also reported in studies that explicitly used this form of superficial or systemic signal control. However, the subset is too small and methodologically heterogeneous to support a formal sensitivity analysis, and this observation does not establish that systemic physiological or motion-related confounding was eliminated.

Taken together, methodological heterogeneity occurred at both the instrumentation and analysis levels, limiting direct comparison of hemodynamic findings across studies and making the current evidence more suitable for descriptive methodological mapping than for quantitative cross-study synthesis.

\subsection{Study-Level Synthesis}
\label{sec:detailed}

The reviewed studies were grouped into two categories: comparative studies, which examined differences across instructional formats, interaction modalities, and multimedia delivery platforms; and cognitive response studies, which focused on neural and expertise-related states when using game-integrated learning systems. 

\subsubsection{Comparative Studies}

Comparative studies used fNIRS to examine task-related cortical
hemodynamic responses across different learning conditions, interaction
modalities, and learner characteristics, often alongside behavioral,
performance, or other measures used by the primary studies to examine
cognitive processes.
Across the comparative studies, the prefrontal cortex was the most
frequently examined region, although several studies also measured
parietal, motor, or other cortical regions. Reported hemodynamic
responses varied across experimental conditions, tasks, and participant
groups.

\paragraph{Platform-Related Findings and the Brain--Learning Relationship}

Several studies reported both neurophysiological and behavioral or
learning-related outcomes. \citet{LAMB201814} reported a significant
positive correlation ($r = 0.89$) between cortical hemodynamic responses
and learning outcome across the examined conditions and also measured
engagement and learning outcomes separately. In contrast,
\citet{de2023engaging} reported higher frontal hemodynamic responses in
the game condition without a corresponding improvement in task
performance relative to the non-game condition.
\citet{Shayla2024StemC} reported a 47.74\% within-group increase in the learning-outcome score in the game group alongside higher cortical hemodynamic responses, although the between-group learning difference was not statistically significant ($N=12$). \citet{Cakir2015AnOptical} reported
decreased responses across the 16 measured prefrontal optodes following
game-based mathematics practice. The primary study interpreted this
decrease in relation to reduced cognitive effort following practice;
the study involved adult participants rather than the target
middle-school population.


In a direct comparison within the same study,
\citet{Ge2021EffectsTwogames} reported primarily prefrontal responses
during mobile interaction and additional motor-cortex responses during
the VR condition. \citet{sharmin2024fnirs} reported higher cortical
hemodynamic responses with stylus input than with hand input, while
performance was similar between the input conditions. These findings
describe condition-specific differences within the respective studies
and do not establish general effects of platform type.
\citet{Lingelbach2023TowardsUserAware} integrated an fNIRS-based
brain--computer interface into a VR learning environment across multiple
sessions. Right dorsolateral prefrontal responses differentiated the
high- and low-working-memory-load conditions during the first session,
whereas this distinction diminished in later sessions as task difficulty
increased.


Several comparative studies examined hemodynamic responses under different task demands. \citet{Shibata2019ImplicitDialogue} reported differences between hard and easy memory-task conditions.
\citet{LANDOWSKA2024103206} reported lower lateral prefrontal hemodynamic responses in the adaptive-training group relative to the non-adaptive control group for a high-demand working-memory condition after training, with part of this difference maintained at follow-up. The primary study interpreted these hemodynamic differences as reflecting reduced mental workload. However, the adaptive intervention did not produce a significant improvement in Starry Night working-memory accuracy relative to the control condition; reported improvements in localization performance occurred across sessions in both groups. Therefore, the study does not establish that adaptive difficulty simultaneously reduced cognitive load and improved task performance.

\paragraph{Methodological and Ecological Boundaries of Comparative Studies}

Across three studies from the same research group, lower prefrontal
responses were reported with greater task experience or across training.
The studies examined VR-based Basic Life Support training
\citep{Aksoy2019PerformanceMonitoring,Aksoy2019Correlation} and
intravenous catheterization simulation
\citep{Aksoy2020AssessingIntravenous}. The authors reported differences
in left prefrontal oxygenation between participants with different
levels of experience and changes in these responses alongside task
familiarity or performance.

\citet{LANDOWSKA2024103206} extended this line of investigation 
with a substantially larger sample ($N = 85$) in an AR setting, compared 
with the smaller samples examined in the Aksoy studies. Across these studies, lower prefrontal hemodynamic responses were reported in association with training, task experience, or experimental condition, but the accompanying behavioral findings were not consistent across studies. In \citet{LANDOWSKA2024103206}, the adaptive-training group showed lower prefrontal responses in a high-demand working-memory condition, whereas intervention-specific improvement in working-memory accuracy was not demonstrated. 
None of these studies separately established that the observed changes
in cortical hemodynamic responses represented learning gains. This
limits interpretation of the neural changes independently of other
possible task- or session-related processes.

In studies involving social or collaborative conditions,
\citet{GUI2025105266} reported inter-brain synchrony associated with
knowledge retention ($r=0.688$), while
\citet{Lei2023EmotionalResponse} reported condition-specific prefrontal
differences during cooperative and competitive play without a
corresponding difference in memory accuracy. These findings describe
associations and condition-specific differences rather than causal
relationships.
Most comparative studies were conducted in controlled settings.
\citet{WANG2025272} conducted a study in a school setting and reported an
association between post-intervention intraparietal sulcus responses
and shorter arithmetic reaction times, although no significant overall
behavioral improvement was reported.

\subsubsection{Cognitive Response Studies}

These studies used fNIRS to characterize task-related cortical
hemodynamic responses in GILS contexts without primarily focusing on
between-group comparisons. Prefrontal responses were frequently
reported, while additional cortical regions were examined depending on
the task and measurement configuration.

\paragraph{Task Demands and Learner-State Classification}

Three studies examined cortical hemodynamic responses in relation to
task demands or task conditions. \citet{Sevecenko2022NeuralCorrelates}
reported prefrontal responses that varied across phases of an emergency
simulation and were examined alongside self-reported workload.
\citet{Ninaus2014PotentialUse} reported frontal and parietal differences
between learning-oriented gameplay and random button-pressing conditions.
\citet{BakeMoyeTuck2018ac} reported responses in a region associated in
prior literature with numerical processing during touchscreen
mathematics-app use, although learning outcomes were not measured.
These studies related hemodynamic responses to cognitive or task
constructs, but the fNIRS measures themselves did not directly measure
mental effort, engagement, or learning.

Two studies used fNIRS-derived features for participant-state
classification. \citet{brainsci11010106} combined fNIRS and facial
expression data to distinguish novice, intermediate, and expert players,
while \citet{ZHANG2021102547} used a drag-and-drop game task to classify
mild cognitive impairment, reporting accuracy up to 90.1\%.
The former involved observation of gameplay videos rather than active
gameplay, whereas the latter used a single laboratory session without
follow-up.

\paragraph{Methodological Limits and Ecological Boundaries of Cognitive Response Studies}

Across these cognitive response studies, learning outcomes were not
consistently assessed alongside the neurophysiological measures.
\citet{Ninaus2014PotentialUse} interpreted fronto-parietal responses in
relation to rule learning, but the neurophysiological pilot included
four participants and the fNIRS data were not subjected to inferential
statistical testing. \citet{BakeMoyeTuck2018ac} related the observed
responses to prior neuroscience findings on numerical processing and
described the study as exploratory and hypothesis-generating.
\citet{Sevecenko2022NeuralCorrelates} examined prefrontal responses
alongside self-reported workload while noting that performance
differences could not be attributed to cognitive load alone.

The two classification studies addressed different populations and
tasks. \citet{brainsci11010106} classified expertise defined through
game experience while participants observed gameplay videos rather than
playing actively. \citet{ZHANG2021102547} examined mild cognitive
impairment in adults aged 50--81 using a single-session drag-and-drop
task. Neither study directly assessed learning gain within a
game-integrated learning intervention.
Across the five cognitive response studies, analyzed sample sizes ranged
from $N=4$ to $N=37$, and the studies were conducted in controlled
settings. The populations, tasks, and outcomes also differed across
studies. These characteristics are reported descriptively rather than
being used to assign methodological-quality ratings.

\section{Discussion}
\label{sec:Discussion}
%
\subsection{\texorpdfstring{$RQ_1$}{RQ1}: Task-Related Hemodynamic Responses Across Game-Integrated Learning Contexts}

In answer to $RQ_1$, task-related cortical hemodynamic responses were
most frequently reported from the prefrontal cortex, although other
regions were examined depending on the task and measurement configuration.
Individual studies reported differences in hemodynamic responses across
experimental conditions or stages of training. For example, several
studies reported reduced prefrontal hemodynamic responses across training
or task conditions
\citep{Aksoy2019PerformanceMonitoring,
LANDOWSKA2024103206}. While the primary studies interpreted some of these
changes in relation to constructs such as cognitive load, skill
development, or cognitive efficiency, we treat these as study-specific
interpretations rather than direct measures of those constructs.
Increased or decreased hemodynamic responses may have multiple
interpretations depending on the task and experimental context, and such
interpretations are more informative when supported by complementary
behavioral, learning, or subjective measures.

Differences observed across separate studies should not be interpreted as
direct platform effects. The included studies differed in learning
content, participant characteristics, task demands, study duration,
fNIRS instrumentation, optode placement, cortical coverage, and
analytical procedures. We therefore distinguish findings from studies
that directly compared conditions within the same experimental design
from differences observed across studies that were not directly
comparable. In addition, the predominance of prefrontal findings partly
reflects the frequent measurement of this region across the reviewed
studies and should not be interpreted as evidence that the prefrontal
cortex is necessarily the most relevant region across game-integrated
learning contexts.

A small number of studies measured both brain activation and learning outcomes together and found correlations between them \citep{LAMB201814, GUI2025105266}. 
The absence of a separate learning measure in half of the corpus limits
the extent to which task-related hemodynamic responses can be related
to learning outcomes. However, because the primary studies generally
did not report their reasons for omitting learning measures, we do not
attribute this pattern to resource constraints, design preferences, or
field norms.
Taken together, the evidence does not establish that specific multimedia platforms are better suited to particular learning outcomes. Individual studies reported associations involving motor interaction, collaboration, performance, or learning, but cross-study comparisons cannot isolate platform effects from methodological differences, and no included study established a causal relationship between fNIRS-derived hemodynamic responses and learning outcomes.

CTML provides a useful theoretical context for considering how multimedia
design may shape cognitive processing. However, the studies in this review
were not selected or coded according to specific CTML principles, and most
did not directly manipulate or test predictions such as the modality or
redundancy effects. Therefore, the available evidence does not support a
systematic principle-by-principle mapping between CTML predictions and
fNIRS responses. The reviewed studies instead show that task-related
cortical hemodynamic responses vary across different multimedia and
game-integrated learning contexts. Future studies could explicitly
manipulate CTML-relevant design features while measuring both
neurophysiological responses and learning outcomes to examine whether
specific multimedia principles are associated with reproducible
hemodynamic patterns.

Combining neurophysiological measures with behavioral, learning, or
subjective outcomes can strengthen interpretation when these measures
are relevant to the research question. Not every study requires all four
types of measures. However, complementary measures are particularly
important when cortical hemodynamic responses are interpreted in relation
to constructs such as cognitive load, engagement, or learning, because
fNIRS does not directly measure these constructs. Future research should
also examine whether fNIRS provides information that meaningfully extends
what can be obtained from less intrusive measures such as behavioral
performance, questionnaires, or interaction logs. Demonstrating such
added value is important when considering the additional cost,
complexity, and measurement burden of fNIRS in educational settings.

\subsection{\texorpdfstring{$RQ_2$}{RQ2}: Methodological Standardization and Ecological Validity}

\textbf{Methodological heterogeneity.} The methodological heterogeneity of the reviewed fNIRS studies also
limits direct comparison of cortical hemodynamic findings across the
corpus. Devices differed in channel configuration and cortical coverage,
while preprocessing and artifact-handling procedures were not uniform.
In particular, explicit approaches for separating cortical signals from
superficial or systemic physiological contributions, such as
short-channel regression, were reported in only a subset of studies.
Consequently, reported differences in oxygenated or deoxygenated hemoglobin should not be assumed
to represent directly comparable estimates of neural activity across
studies. We therefore interpret cross-study hemodynamic patterns
descriptively and consider differences in instrumentation, cortical
coverage, preprocessing, and signal-quality control when discussing
the evidence. Reporting of VR-related discomfort was inconsistent across the six studies involving immersive VR or AR configurations. Some studies formally assessed or discussed cybersickness, discomfort, or related screening criteria, whereas others did not report this information consistently. Although Aksoy et al. acknowledged dizziness and headset-related discomfort as potential limitations, the number of participants experiencing these symptoms or excluded specifically because of them was not reported~\cite{Aksoy2019Correlation}.

\textbf{Sample size, power justification, and ecological setting.}
Sample sizes varied substantially across the corpus. Four of the 20
included studies explicitly reported a formal power analysis
\citep{LAMB201814, Lei2023EmotionalResponse, WANG2025272,
sharmin2024fnirs}. For the remaining studies, a formal power analysis
was not identified in the primary report. The absence of a reported
power analysis should not be interpreted as evidence that no sample-size
planning was undertaken, and sample size alone was not used as an
indicator of methodological quality. Most studies were conducted in
controlled settings, whereas \citet{WANG2025272} examined a game-based
mathematics intervention in a school setting. Further research in
authentic educational settings would help establish whether reported
findings generalize beyond controlled study contexts.

\textbf{Real-time state estimation and adaptation.}
Here, we use ``real-time'' to refer to fNIRS-derived state estimation
performed during the ongoing experimental session, rather than analysis
conducted between separate sessions. Shibata et
al.~\cite{Shibata2019ImplicitDialogue} demonstrated within-session,
participant-specific fNIRS-based classification of busyness. However,
in the proof-of-concept experiment, known task difficulty was used as a
proxy for busyness to select the interruption strategy, while the
fNIRS-derived estimates were generated in real time and logged for
evaluation. Thus, the study provides preliminary evidence for the
technical feasibility of within-session fNIRS-based state
classification, rather than evidence that a closed-loop fNIRS-driven
adaptive learning system improved learning outcomes.

\textbf{fNIRS as a tool for GILS.} Some studies in this corpus reported hardware challenges in movement-heavy and immersive GILS contexts, including signal quality issues and user discomfort. Recent advances in wearable fNIRS platforms show progress in motion tolerance and user comfort \citep{Upadhyay2025}. As wearable systems continue to develop, fNIRS may provide a useful
research approach for examining task-related cortical hemodynamic
responses in multimedia learning environments. Its added value relative
to less intrusive measures, however, requires further validation.

\subsection{Future Directions} \label{sec:future_direction}

This review identifies four directions for future fNIRS-based GILS research that may also inform multimedia learning system design more broadly.

\textbf{Scope 1: Understanding the brain--learning relationship.} 
Depending on the research question, combining neurophysiological measures with behavioral, learning, or subjective measures may support a broader interpretation of task-related responses. Pre- and post-assessments may help interpret neural activity alongside learning outcomes. Longitudinal designs could help researchers examine how these relationships develop over time and across learning contexts.

\textbf{Scope 2: Improving comparability and ecological validity.} More consistent reporting of optode placement, preprocessing procedures, signal selection, and effect sizes would allow findings to be compared more easily across studies. 

Reporting an a priori sample-size or power justification, where
appropriate to the study design, would improve transparency regarding
whether the planned sample was adequate for the intended analysis.

Conducting research in authentic classrooms and learning environments may also produce evidence that is more relevant to real-world educational practice.


\textbf{Scope 3: Evaluating real-time neuroadaptive systems.}
The reviewed evidence does not yet establish that fNIRS-driven adaptation
improves learning. Future work should evaluate the generalizability of
within-session state estimation across participants, tasks, and learning
contexts, including the effects of individual calibration, processing
latency, and signal quality. Neuroadaptive approaches should also be
compared with less intrusive indicators, such as behavioral performance
or interaction logs, to examine whether fNIRS provides meaningful
added value. Most importantly, future studies should evaluate whether
adaptations based on these signals produce measurable learning benefits.
Classroom deployment would additionally require attention to neuroprivacy,
informed consent, algorithmic bias, student autonomy, and the collection,
storage, and use of students' brain data.

\textbf{Scope 4: Expanding learner diversity.} Most studies focused on healthy adults or university students, while only a few examined children, learners with cognitive variability, or individuals with mild cognitive impairment \citep{BakeMoyeTuck2018ac, WANG2025272, ZHANG2021102547}. Expanding research across a wider range of learner populations would help build a more inclusive understanding of how different learners respond to multimedia and game-based learning environments.

\subsection{Limitations of This Review}

The final corpus consisted of 20 studies, reflecting the relatively early stage of research at the intersection of fNIRS and game-integrated learning. While this corpus was sufficient to identify common methodological and thematic patterns, the findings are intended to characterize the current landscape of the field rather than provide statistically generalizable conclusions.
The studies used different age groups, different fNIRS devices, different game types, and different outcome measures. This makes it hard to draw deep conclusions within any single topic area. Some studies could also reasonably fit into more than one category in the classification framework. The final assignment involved judgment from the review team.
This review included only English-language publications, so relevant studies published in other languages may have been missed. The review was also restricted to published empirical studies; although eligible conference publications were included, theses, dissertations, preprints, and other unpublished or non-indexed work were not represented in the final corpus. Relevant evidence from these sources may therefore have been missed. In addition, studies with null or non-significant findings may be less likely to appear in the published literature, which may result in publication bias and a more positive representation of fNIRS in game-integrated learning than the full body of evidence supports. Most records (1,433 of 1,563) were retrieved from Google Scholar. Its limited filtering options required extensive manual screening, which may have introduced variability in the selection process.

Additional limitations concern the scope and review process. We did not conduct a formal risk-of-bias assessment using a validated appraisal instrument; instead, we report selected study-level methodological characteristics to support transparent interpretation of the evidence. The search did not include databases such as Scopus, Web of Science, ERIC, or PsycINFO. In addition, the iterative search expressions and Google Scholar's search interface may limit exact replication of the search process. The boundaries of game-integrated learning also required judgment for a small number of included studies. Some studies examined contexts near the boundary of this scope, including video-based game observation or cognitive assessment, which increases the heterogeneity of the corpus. Finally, initial data extraction was conducted by one researcher. Although additional reviewers subsequently cross-checked the extracted information against the primary studies and any identified extraction errors were corrected, the use of a single initial extractor remains a limitation.

\section{Conclusion} \label{sec:Conclusion}
This scoping review synthesized 20 empirical studies using fNIRS in game-integrated learning systems (GILS), examining their educational contexts, neurophysiological findings, and methodological approaches. Regarding $RQ_1$, the reviewed studies reported task-related cortical hemodynamic responses across diverse learning platforms and tasks, most frequently in prefrontal regions, with some studies reporting associations with behavioral or learning outcomes. However, these responses do not directly establish learning, engagement, or cognitive load, and differences observed across separate studies cannot be attributed to platform design alone.
Regarding $RQ_2$, the literature showed substantial variation in study design, sample size, fNIRS configuration, cortical coverage, preprocessing, outcome measures, and ecological settings. These differences limit cross-study comparisons and highlight the need for more consistent reporting, stronger methodological validation, and research in authentic educational settings. Evidence supporting real-time fNIRS-based neuroadaptive learning remains preliminary. Future research should establish whether neurophysiological measures provide added value beyond behavioral and other less intrusive measures, whether state estimates generalize across participants and learning contexts, and whether adaptations based on these signals produce measurable learning benefits while addressing the ethical implications of collecting students’ brain data.
Methodologically, this review contributes a structured mapping of learning contexts, multimedia delivery platforms, outcome measures, study designs, fNIRS instrumentation, preprocessing approaches, and analytical procedures across the available GILS literature, providing a transparent basis for identifying methodological gaps and priorities for future research. fNIRS may serve as a complementary research tool for examining task-related cortical hemodynamic responses during interaction with game-integrated multimedia environments, but its value for adaptive learning will depend on validated links between these signals and meaningful learning outcomes.

\section*{Statements and Declarations}
\subsection*{Availability of data and materials}
The datasets generated and analyzed during the current study are available from the corresponding author on reasonable request. The list of included studies is provided in ~\autoref{tab:listed paper}.
\subsection*{Competing interests}
The authors declare that they have no competing interests.
\subsection*{Funding}
This work was supported by the National Science Foundation (NSF) under grants \#2222661-2222663, \#2321274, and \#2426003. 
 The authors also acknowledge funding from the Institute for Engineering-Driven Health at the University of Delaware and the National Science Foundation Accelerating Research Translation program under award number \#2331440. Any opinions, findings, and conclusions expressed in this material are those of the authors and do not necessarily reflect the views of the funding agency.
  
\subsection*{Authors' contributions}

\textit{S. Sharmin:} conceptualization, methodology, data curation, screening, result analysis, original draft writing, review, and editing. 
\textit{G. Lucero-Palacios:} data curation, screening, review, and editing. 
\textit{B. Kiafar:} data curation, screening, review, and editing. 
\textit{M.F. Abrar:}  data curation, screening, review, and editing. 
\textit{M. Al-Ratrout:} screening, review, and editing. 
\textit{A. Raikwar:} methodology, review, and editing. 
\textit{R.L. Barmaki:} (Supervision) conceptualization, methodology, review, and editing. 

\subsection*{Acknowledgments}
The authors thank all contributors who assisted with study screening and data curation.
\bibliography{sn-bibliography}
\appendix
\section*{Appendix 1: Complete Search Strategy}\label{appendix 1}

\begin{table*}[h]
\centering
\caption{Databases, search dates, search queries, and records retrieved
across the three search phases.}
\label{tab:supp_search_strategy}
\small
\setlength{\tabcolsep}{5pt}
\renewcommand{\arraystretch}{1.2}

\begin{tabular}{
p{2.5cm}
p{3.5cm}
p{2.0cm}
r
r
r
r}
\hline
\textbf{Database / Source} &
\textbf{Search date(s)} &
\textbf{Queries} &
\textbf{2024} &
\textbf{2025} &
\textbf{2026} &
\textbf{Total} \\
\hline

Google Scholar &
June 13--16, 2024; June 2, 2025; April 1, 2026 &
Q1--Q5 &
888 & 246 & 299 & 1,433 \\

ACM Digital Library &
June 13--16, 2024; June 2, 2025; April 1, 2026 &
Q1--Q5 &
35 & 29 & 1 & 65 \\

IEEE Xplore &
June 13--16, 2024; June 2, 2025; April 1, 2026 &
Q1--Q5 &
6 & 0 & 0 & 6 \\

PubMed &
June 13--16, 2024; June 2, 2025; April 1, 2026 &
Q1--Q5 &
1 & 0 & 0 & 1 \\

ScienceDirect &
June 13--16, 2024; June 2, 2025; April 1, 2026 &
Q1--Q5 &
27 & 27 & 4 & 58 \\

\hline
\textbf{All sources} &
&
&
\textbf{957} &
\textbf{302} &
\textbf{304} &
\textbf{1,563} \\
\hline
\end{tabular}
\end{table*}

\noindent
\textbf{Complete search queries:}

\begin{enumerate}
    \item[Q1.] ``game-based learning'' and fNIRS
    \item[Q2.] ``educational game'' and fNIRS
    \item[Q3.] ``serious game'' and fNIRS
    \item[Q4.] (``serious game'' or ``educational game'' or
    ``game-based learning'') and ``fNIRS''
    \item[Q5.] gamification and fNIRS
\end{enumerate}

\noindent
\textit{Note.}
The same five search queries were used across the three search phases. For Google Scholar, all result pages made available by the platform at the time of each search were manually reviewed rather than applying a predefined page or result cutoff. No automated scraping was used. Records were manually saved using Google Scholar’s ``Save'' function and subsequently exported from the Google Scholar Library as CSV files for screening. Google Scholar records were not assumed to be mutually exclusive from records retrieved through the other sources; duplicate records were removed during screening.
\newpage
\section*{Appendix 2: Screening and Exclusion Summary}

\begin{table*}[h]
\centering
\caption{Screening and exclusion summary across the three search phases.}
\label{tab:supp_screening}
\small
\setlength{\tabcolsep}{5pt}
\renewcommand{\arraystretch}{1.3}

\begin{tabular}{
p{1.2cm}
p{1.1cm}
p{1.45cm}
p{1.3cm}
p{1.8cm}
p{1.0cm}
p{1.6cm}
p{1.0cm}
}
\hline
\textbf{Search phase} &
\textbf{Records} &
\textbf{Duplicates} &
\textbf{Non-English} &
\textbf{Other early exclusions} &
\textbf{No fNIRS} &
\textbf{Final exclusions} &
\textbf{Included} \\
\hline

2024 & 957 & 25 & 87 & 148 & 510 & 170 & 17 \\

2025 & 302 & 15 & 20 & 226 & 14 & 25 & 2 \\

2026 & 304 & 5 & 14 & 60 & 195 & 29 & 1 \\

\hline
\textbf{Total} &
\textbf{1,563} &
\textbf{45} &
\textbf{121} &
\textbf{434} &
\textbf{719} &
\textbf{224} &
\textbf{20} \\
\hline

\end{tabular}

\vspace{2mm}

\begin{minipage}{0.96\textwidth}
\footnotesize
\textit{Note.} ``Other early exclusions'' includes review articles,
books, chapters, abstracts, theses, and other records considered
ineligible or outside the scope of the review. ``Final exclusions''
refers to records excluded at the subsequent eligibility stage because
they did not meet the game-integrated learning context criterion.
\end{minipage}

\end{table*}
\end{document}